\documentclass[aps,prb,twocolumn,showkeys,amsmath,amssymb,longbibliography,floatfix,superscriptaddress,groupedaddress]{revtex4-1}
\usepackage[T1]{fontenc}
\usepackage[utf8]{inputenc} % Ensure compatibility with UTF-8
\usepackage{lmodern}        % Improved font rendering

\usepackage{amsmath,amssymb,amsfonts,bm,dsfont,units}
\usepackage{braket}         % For quantum states
\usepackage{wasysym}        % Additional symbols

\usepackage{array,makecell,multirow,booktabs,hhline}

\usepackage{graphicx}
\usepackage{float}          % Float placement control
\usepackage[caption=false]{subfig} % For subfi.gures (compatible with RevTeX)

\usepackage{xcolor}
\usepackage{colortbl}       % Colored table cells
\usepackage{tikz}           % Drawing and diagrams

\usepackage{hyperref}
\hypersetup{
    colorlinks=true,
    citecolor=blue,
    linkcolor=blue,
    urlcolor=blue,
    pdfstartview=FitH
}

\usepackage{appendix}
\usepackage{diagbox}        % Diagonal cell split in tables
\usepackage{comment}
\usepackage[normalem]{ulem} % Strikeout text with \sout{}

\newcommand{\vect}[1]{\boldsymbol{#1}} % Vector notation
\begin{document}

\title{Universal charge conductance at Abelian--non-Abelian quantum Hall interfaces}
\author{Misha Yutushui}
\affiliation{Department of Condensed Matter Physics, Weizmann Institute of Science, Rehovot 76100, Israel} 
\author{Ady Stern}
\affiliation{Department of Condensed Matter Physics, Weizmann Institute of Science, Rehovot 76100, Israel} 
\author{David F. Mross}
\affiliation{Department of Condensed Matter Physics, Weizmann Institute of Science, Rehovot 76100, Israel} 
\begin{abstract}
    Multiple topologically distinct quantum Hall phases can occur at the same Landau level filling factor. It is a major challenge to distinguish between these phases as they only differ by the neutral modes, which do not affect the charge conductance in conventional geometries. We show that the neutral sector can be determined with coherent charge conductance in a $\pi$-shaped geometry that interfaces three different filling factors. Specifically, non-Abelian paired states at a half-filled Landau level and the anti-Read-Rezayi state can be identified. Interestingly, for interfaces between paired states and Jain states, the electric current in the $\pi$ geometry behaves as if pairs of neutral Majoranas edge modes were charge modes of Jain states.
\end{abstract}

\date{\today}
\maketitle

The fractional quantum Hall effect~\cite{Tsui_fqh_1982,Laughlin_fqh_1983,Haldane_fqh_1983,Halperin_fqh_1983} is the primary platform for realizing topological order. The fractional quantum Hall states that feature non-Abelian anyons~\cite{Wen_Non-Abelian_1991,Moore_nonabelions_1991,Read_paired_2000,Stern_Probe_Non_Abelian_2006,nayak_non-abelian_2008,Stern_non_Abelian_2010} are especially interesting due to their promise of fault-tolerant quantum computing.~\cite{Freedman_modular_functor_2000,Bonesteel_Braid_Topologies_2005,Hormozi_Topological_quantum_2007,nayak_non-abelian_2008,Stern_non_Abelian_2010} The most well-known candidate for non-Abelian states occurs at half-filling of the first excited Landau level in GaAs, i.e., at $\nu=\frac{5}{2}$.~\cite{Willett_observation_1987} Similar plateaus were also observed in monolayer,~\cite{Kim_Even_Denominator_f_wave_2019} bilayer~\cite{Ki_bilyaer_graphene_2014,Kim_bilayer_graphene_2015,Li_bilayer_graphene_2017,Zibrov_Tunable_bilayer_graphene_2017,Assouline_Energy_Gap_bilayer_graphene_2024,Kumar_Quarter_2024} and trilayer~\cite{chen_tunable_2023} graphene and other heterostructures.~\cite{Falson_Zno_2015,Falson_Zno_2018} These putative non-Abelian states can be fruitfully understood as paired states of composite fermions~\cite{Read_paired_2000}. The pairing channels identified by numerics~\cite{Morf_transition_1998,Rezayi_incompressible_2000,Wojs_landau_level_2010,Storni_fractional_2010,Feiguin_spin_2009,Peterson_Finite_Layer_Thickness_2008} point towards a Moore-Read state~\cite{Moore_nonabelions_1991} with a single downstream Majorana mode at the edge ($\ell=1$), or its particle-hole conjugate, the anti-Pfaffian state~\cite{Lee_particle_hole_2007,Levin_particle_hole_2007} with three upstream Majorana modes ($\ell=-3$).~\cite{Rezayi_breaking_2011,Pakrouski_phase_2015,Rezayi_Landau_2017}
 
Thermal Hall conductance~\cite{Banerjee_observation_2018,Dutta_Isolated_2022,Paul_Thermal_2024,Melcer_Heat_2024} and noise measurements~\cite{Dutta_novel_2022} in GaAs indicate a different topological order, the particle-hole symmetric Pfaffian (PH-Pfaffian) phase ($\ell=-1$).~\cite{Son_is_2015,Bonderson_time-reversal_2013,Chen_symmetry_2014}. In contrast, in bilayer graphene, the daughter states~\cite{Levin_collective_2009,Yutushui_daughters_2024,Zheltonozhskii_daughters_2024,zhang_hierarchy_2024} observed near even-denominator states suggest Moore-Read or anti-Pfaffian orders. Alternative charge and noise probes of topological orders at half filling were suggested ~\cite{Manna_Full_Classification_2022,Yutushui_Identifying_2022,Park_Fingerprints_2024,Yutushui_Identifying_2023} but were thus far not realized.  

\begin{figure}[t]
 \centering
\includegraphics[width=0.45\linewidth]{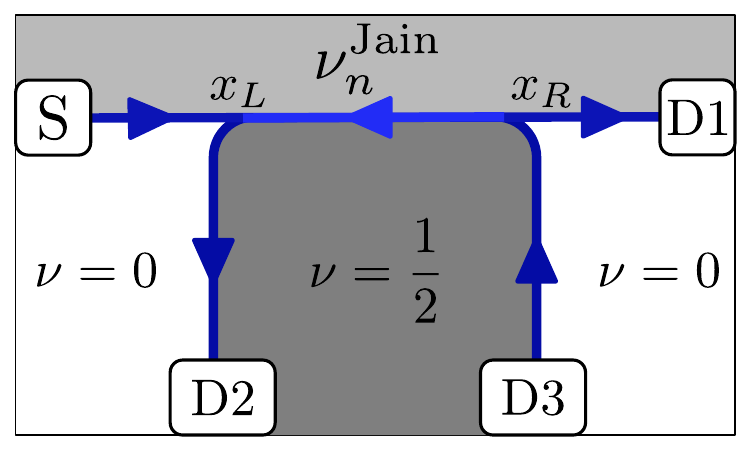}
\includegraphics[width=0.45\linewidth]{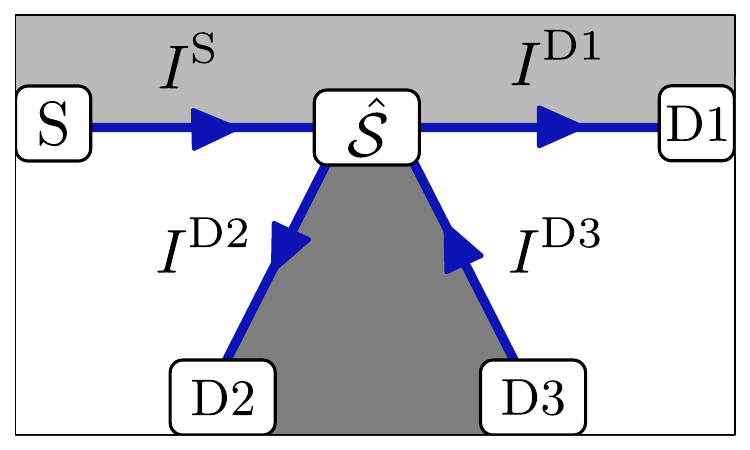}
  \caption{Left panel: The $\pi$-geometry consists of a region with a half-filled state interfaced along the constriction $x\in[x_L,x_R]$ with a Jain state. All leads are connected to contacts. The source S contact is kept at voltage $V$, while the drains D1, D2, and D3 are grounded. Right panel: The constriction $x\in[x_L,x_R]$ can be effectively replaced by a scattering matrix $\hat{\cal S}$ that sets boundary conditions between the currents $I^\text{S}_a$, $I^\text{D1}_a$, $I^\text{D2}_0$, and $I^\text{D3}_0$.}
 \label{fig.1} 
 \end{figure}

In this letter, we show that charge conductance in $\pi$-geometry~\cite{Yutushui_Identifying_2022} in Fig.~\ref{fig.1}~(a) can be used to identify \textit{any} non-Abelian pairing channel. Moreover, the $\pi$-geometry can confirm the presence of the anti-Read-Rezayi state at $\nu=\frac{12}{5}$ featuring Fibonacci anyons praised for their application to universal fault-tolerant quantum computing.~\cite{Freedman_modular_functor_2000,Bonesteel_Braid_Topologies_2005,Hormozi_Topological_quantum_2007,nayak_non-abelian_2008,Stern_non_Abelian_2010} We demonstrate that when the non-Abelian state in question, e.g., the $\nu=\frac{1}{2}$ in Fig.~\ref{fig.1}~(a), is interfaced with suitable reference states, the coherent charge conductance depends on the topological order of the non-Abelian state, and thus can be used to identify the latter. Different conductance values are the result of the localization of counter-propagating modes in the interface, which depends on the neutral modes distinguishing different non-Abelian phases at the same filling factor. Finally, we show that this setup leads to a new transport regime where the charge conductance takes a quantized value that is distinct from other transport regimes.

We use well-known Jain states as reference states to probe the non-Abelian state under consideration; see Fig.~
\ref{fig.1}(a). At low temperatures and voltages, the relatively short middle region (constriction) $x\in [x_L,x_R]$ can be viewed as a point-like scatterer $\hat{\cal S}$; see Fig.~\ref{fig.1}~(b). The current emanating from the source contact S is partitioned among the leads flowing to the drain contacts D1 and D2. The two-terminal conductance $G$ between the source S kept at voltage $V$ and the drain D1 strongly depends on the equilibration regime in the constriction. If the constriction's length $L=x_R-x_L$ is sufficient to achieve charge equilibrium, the conductance is given by the difference of filling factors $\delta \nu = \nu_\text{Paired}-\nu_\text{Jain}$, i.e., $G_\text{equilibrated}=\text{min}(-\delta \nu,0)$. In the opposite limit of a short constriction, the random tunneling is inefficient; all modes propagate unimpeded, and the conductance is ballistic $G_\text{ballistic}=\nu_\text{Jain}$. The conductance in both of these regimes is insensitive to the topological order of the non-Abelian states and depends only on the filling factors of the Jain and paired states.

We now study the coherent effects of disorder on transport and reveal an intermediate regime that permits the identification of the paired state. This coherent regime is defined by the temperature $T$ being small compared to the energy scale defined by the constriction's length, $T\ll v/L$, where $v$ is a characteristic velocity scale.

The tunneling in the constriction is assisted by disorder and can be a relevant operator, depending on non-universal interactions between the modes. The relevant disorder tunneling terms set the associated length scale $\xi$, at which the tunneling strength becomes of order unity. Below, we collect the disorder effects into a single point by performing a disorder-dependent transformation.~\cite{Protopopov_transport_2_3_2017,Yutushui_Identifying_2022} At this point, the disorder manifests in a set of boundary conditions and residual point-like tunneling terms that depend on the disorder configuration; see Fig.~\ref{fig.1} (b). We further identify the boundary conditions that result in a fixed point theory and assess their stability by the relevance of residual tunneling. 

Two qualitatively different scenarios arise: (i) the edge is topologically stable,~\cite{Haldane_stability_1995} (ii) the edge is topologically unstable, and counter-propagating modes can localize if the edge is sufficiently long $L\gg \xi$ and the corresponding tunneling operator is relevant. We find that in the first case, the only stable fixed point corresponds to trivial boundary conditions, which yield the ballistic conductance value. In the second case, the conductance value satisfies $G_\text{equilibrated}\leq G_\text{coherent}\leq G_\text{ballistic}$. It depends on the localized modes and thus permits identification of the topological order; see Table~\ref{tab.condunctance}. 
\begin{table}[t]
  \centering  
  \caption{The coherent charge conductance $G$ between S and D1 in the setup in Fig.~\ref{fig.1} in the regime $\xi_\text{loc}\ll L\ll L_T$. The cases where the conductance differs from the values in decoupled $L\ll \xi_\text{loc}$ and fully equilibrated regimes $L\gg L_\text{eq}$ are highlighted in boldface. The best-known candidate states are the Moore-Read Pfaffian with $\ell=1$, anti-Pfaffian with $\ell=-3$, PH-Pfaffian with $\ell=-1$, the strong-pairing $K=8$ state with $\ell=0$, 331 with $\ell=2$ and SU(2)$_2$ with $\ell=3$.}
  \renewcommand{\arraystretch}{1.4} 
  \setlength{\tabcolsep}{9pt} 
  \begin{tabular}{c|c c c| c c c}
    \hline \hline 
      $\nu^\text{Jain}_n$  & $\frac{1}{3}$ & $\frac{2}{5}$ & $\frac{3}{7}$ & $\frac{3}{5}$ & $\frac{2}{3}$ & $1$  \\ \hline 
     \diagbox{\vspace{-0.1cm}\hspace{0.3cm}$\ell$}{\vspace{-0.35cm}$n$}& 1 & 2 & 3 & -3 & -2 & -1\\\hline
     $-3,-4$ & $\frac{1}{3}$ & $\frac{2}{5}$ & $\frac{3}{7}$ & $\bf\frac{9}{65}$ & $\frac{1}{6}$ & $\frac{1}{2}$ \\
     $-1,-2$    & $\frac{1}{3}$ & $\frac{2}{5}$ & $\frac{3}{7}$ & $\bf\frac{9}{40}$ & $\bf\frac{4}{15}$ & $\frac{1}{2}$ \\
       $0$ & $\frac{1}{3}$ & $\frac{2}{5}$ & $\frac{3}{7}$ & $\frac{3}{5}$ & $\frac{2}{3}$ & $1$ \\
     $1,2$ & $0$ & $\bf\frac{1}{15}$ & $\bf\frac{2}{21}$ & $\frac{3}{5}$ & $\frac{2}{3}$ & $1$ \\
     $3,4$  &  $0$ & $0$ & $\bf\frac{1}{35}$ & $\frac{3}{5}$ & $\frac{2}{3}$ & $1$ \\
     \hline\hline 
  \end{tabular}
  \label{tab.condunctance}
\end{table}

We derive these results based on the edge theories of Abelian and paired quantum Hall states.~\cite{Wen_edge_1991} Any Abelian topological order can be expressed via a $n\times n$ dimensional integer matrix $K$. Its boundary is governed by the Lagrangian density
\begin{align}\label{eq.LK}
     {\cal L}_{K}  = \frac{1}{4\pi}\sum_{a,b} \partial_x\phi_a(iK_{ab}\partial_\tau-V_{ab}\partial_x)\phi_b\,,
\end{align}
where $V_{ab}$ is a positive definite real symmetric matrix parameterizing velocities and density-density interaction of different edge modes. The electric charge of these modes (in units of the electron charge) is specified by the components of charge vector $\vect{t}$ via $\rho_a =t_a\partial_x\phi_a/2\pi$, which implies the filling factor $\nu=\vect{t}^TK^{-1}\vect{t}$.

The leads from S to $x_L$ and from $x_R$ to D1 are the edges of the Jain state of filling factor $\nu_n=\frac{n}{2n+1}$; see Fig.~\ref{fig.1}. For positive $n$, the leads contain $n$ modes $\phi_a$ with $a\in[1,n]$ governed by Eq.~\eqref{eq.LK} with the $K$-matrix $K^\text{Jain}_n = 2+ \mathbb{I}_n$, where $\mathbb{I}_n $ is the $n\times n$ identity matrix, and the charge vector $t_a=1$. Generalization to a negative $n$ is straightforward; see Appendix~\ref{app.neg_n}. 

 The leads from $x_L$ to D3 and from D2 to $x_R$  are the edges of a non-Abelian paired state and cannot be characterized by a $K$ matrix alone. In addition to Eq.~\eqref{eq.LK} for a single mode $\phi_0$ with $K_\text{Paired}=2$ and $t=1$, the edges contain $\ell$ Majorana fermions, determined by the pairing channel of composite fermions. Their Lagrangian density is
\begin{align}\label{eq.Lgamma}
    {\cal L}_\ell = \sum_{k=1}^{|\ell|} \gamma_k(\partial_\tau-i\overline{v}\partial_x)\gamma_k,
\end{align}
where positive $\ell$ denotes Majorana modes co-propagating with the charge mode (downstream) with average velocity $\overline{v}$. For $\ell<0$, $|\ell|$ Majorana fermions move upstream at the velocity $-|\overline{v}|$. Generically, velocity anisotropy $(v_k-\overline{v})\gamma_k\partial_x\gamma_k$ and interactions between the charge mode and Majorana fermions $\partial_x\phi_a \gamma_i\gamma_j$ are also allowed but are irrelevant in the presence of disorder. 

\begin{figure}[t]
 \centering
\includegraphics[width=0.75\linewidth]{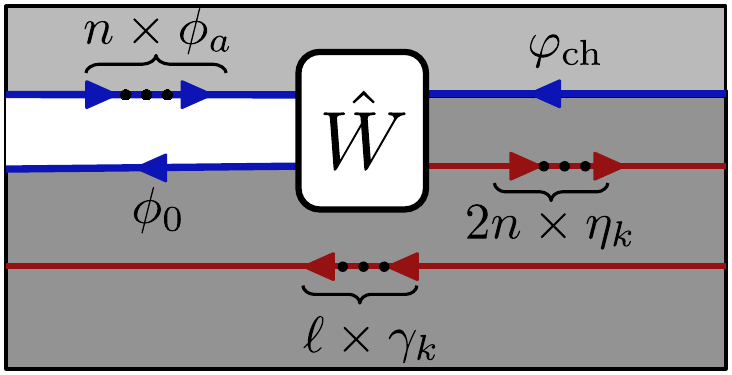}
 \caption{The interface between the Jain state with $n$ charged modes $\phi_a$ and the paired state with a semion mode $\phi_0$ in the standard basis on the left side become after the transformation the charge $\varphi_\text{ch}$ and neutral modes on the right side via Eq.~\eqref{eq.W}. The $n$ neutral boson modes $\varphi_{\sigma,a}$ can be represented as $2n$ Majorana modes $\eta_k$. When $n$ and $\ell$ have equal signs, $m(\ell)$ pairs of $\eta_k$ and $\gamma_k$ Majoranas get localized.}
 \label{fig.2} 
 \end{figure} 

The constriction between $x_L$ and $x_R$ contains counter-propagating edge modes of the paired and Jain states. The interface is described by Eq.~\eqref{eq.LK} with the $n+1$-dimensional $K$-matrix $K=\text{diag}(2,-K^\text{Jain}_n)$ and charge vector $t_a=1$ with $a\in[0,n]$, in addition to $\ell$ downstream Majoranas described by ${\cal L}_{\ell}$ in Eq.~\eqref{eq.Lgamma}. The density-density interaction renders the matrix $V$ non-block-diagonal and introduces new $\partial_x\phi_a \gamma_i\gamma_j$ terms.

Crucial for charge equilibration is electron tunneling between the edges 
\begin{align}\label{eq.L_tun0}
    {\cal L}_\text{tun} =  \sum_{\lambda,\mu} g_{\lambda,\mu}(x) \psi_{e,\lambda}^\dag \psi_{e,\mu}+\text{H.c.},
\end{align}
where $g_{\lambda,\mu}(x)$ is the random tunneling amplitude, and $\lambda,\mu$ enumerate the edge electron operators of the Jain state $\psi_{e,a}\propto \exp i \sum_b K^\text{Jain}_{a,b} \phi_b$ or the paired state $\psi_{e,k}\propto \gamma_k e^{2i\phi_0}$. In the presence of disorder Eq.~\eqref{eq.L_tun0}, the edge renormalizes to a maximally symmetric point, where charge and neutral modes are decoupled, and all neutral modes move at the same velocity; see Refs.~\onlinecite{Moore_Classification_1997,Lee_particle_hole_2007,Levin_particle_hole_2007,Bishara_PH_Read_Rezayi_2008} and Appendix~\ref{app.Fixed points}.

To analyze the constriction, we diagonalize the $K$ matrix by an SL($n$+1,$\mathbb{Z}$) transformation $K'=W^T K W$ with 
\begin{align}\label{eq.W}
            \hspace{-0.3cm}K' =   
    \begin{pmatrix}
     \delta \nu^{-1} &  \begin{matrix} 0& \cdots \end{matrix} \\
    \begin{matrix} 0\\ \vdots \end{matrix} &  \vcenter{\hbox{\scalebox{1.5}{$-\mathbb{I}_n$}}}
    \end{pmatrix}
    \;\;\text{\&}\;\;
    W =   
    \begin{pmatrix}
     2n+1 &  \begin{matrix} -1& \cdots \end{matrix} \\
    \begin{matrix} -2\\ \vdots\end{matrix} &  \vcenter{\hbox{\scalebox{1.5}{$\mathbb{I}_n$}}}
    \end{pmatrix}, 
 \end{align}
and charge vector ${\vect t}'=W^T \vect t = (1,0,\ldots)$, where $\delta \nu^{-1}=4n+2$. The diagonal basis $(\varphi_\text{ch},\varphi_{\sigma,1},\ldots\varphi_{\sigma,n}) = W^{-1}\vect{\phi}$ contains the total charge mode $\varphi_\text{ch} = \vect t^T  \vect \phi$ and $n$ neutral modes $\varphi_{\sigma,a}$. It is convenient to parametrize neutral modes as complex fermions $\psi_{\sigma,a} \propto e^{i\varphi_{\sigma,a}}$ and further represent them via $2n$ Majorana modes $\psi_{\sigma,a}= \eta_{2a-1} + i\eta_{2a}$. In the leads, the density-density interactions in this basis are given by
\begin{align}\label{eq.ch_sigma}
    {\cal L}_{\text{ch}-\sigma}\equiv 2h^\sigma_a \partial_x\varphi_\text{ch} \partial_x\varphi_{\sigma,a} \propto ih^\sigma_a\partial_x\varphi_\text{ch}\eta_{2a-1}\eta_{2a},
\end{align}
where $h_{\sigma,a}=-\frac{4n+2}{4n+1}(v_\text{ch} + v_\sigma)$ corresponds to decoupled paired and Jain edges.

Being neutral, the electron tunneling term Eq.~\eqref{eq.L_tun0} between the Jain and paired edge states amounts to scattering between the $\eta$ Majoranas and the original $\gamma$ Majoranas of Eq.~\eqref{eq.Lgamma}. We encode this disordered free fermion problem by a scattering matrix $\hat{\cal S}$. 

For a more microscopic treatment of disorder in this context, see Refs.~\onlinecite{Kane_Randomness_1994,Lee_particle_hole_2007,Levin_particle_hole_2007,Bishara_PH_Read_Rezayi_2008,Protopopov_transport_2_3_2017,Yutushui_Identifying_2022} and Appendix~\ref{app.Fixed points}. The matrix $\hat{\cal S}$ encodes a boundary condition between the interacting system on either side, i.e., the S, D2 leads to the left, and the D1, D3 leads to its right. The interactions generically renormalize the boundary conditions specified by the scattering matrix. Our strategy is to identify the fixed points of the scattering matrix $\hat{\cal S}$, analyze their stability against weak perturbations, and compute the corresponding conductances.

\begin{figure}[t]
 \centering
\includegraphics[height=3cm]{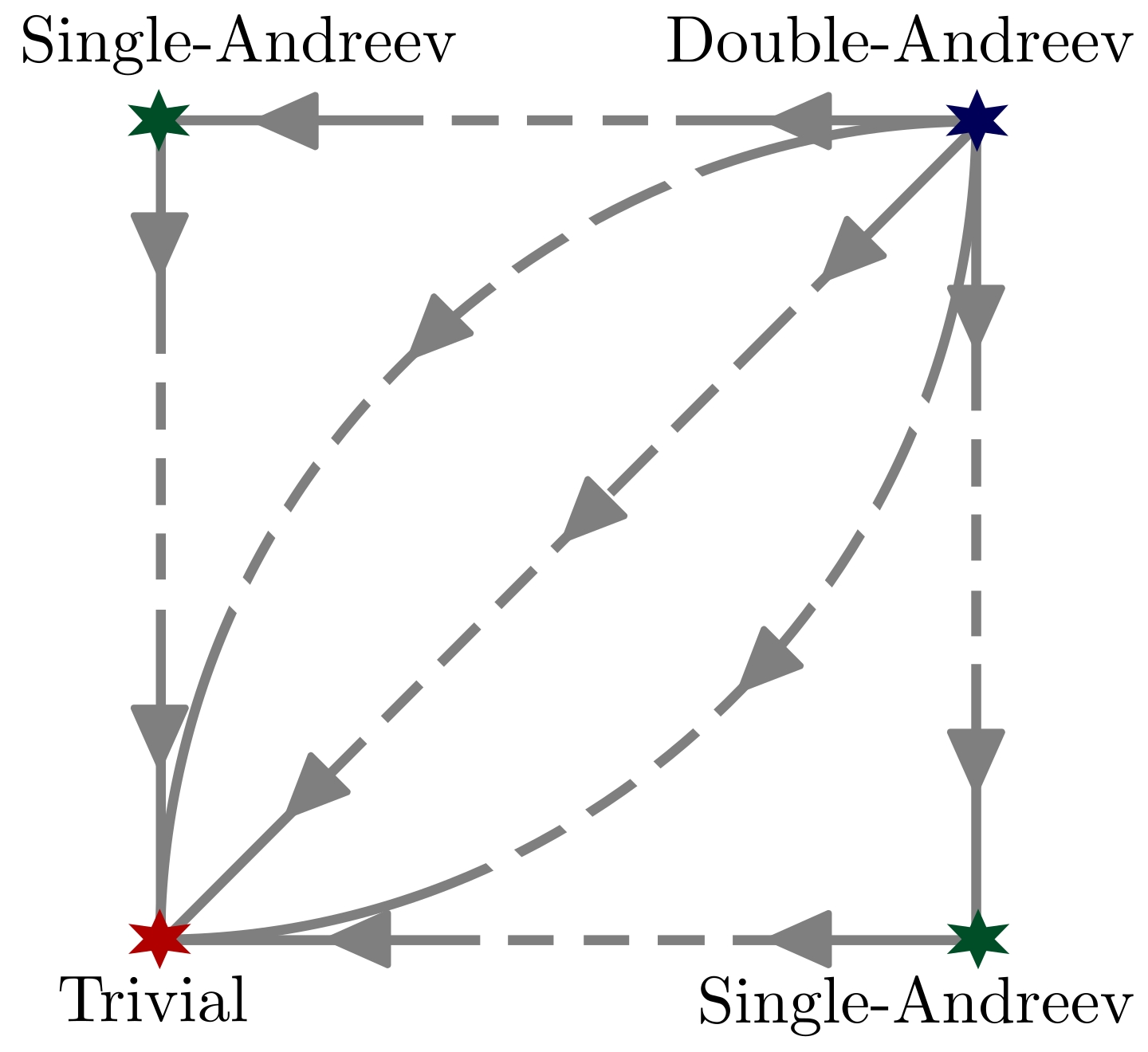}      \includegraphics[height=3cm]{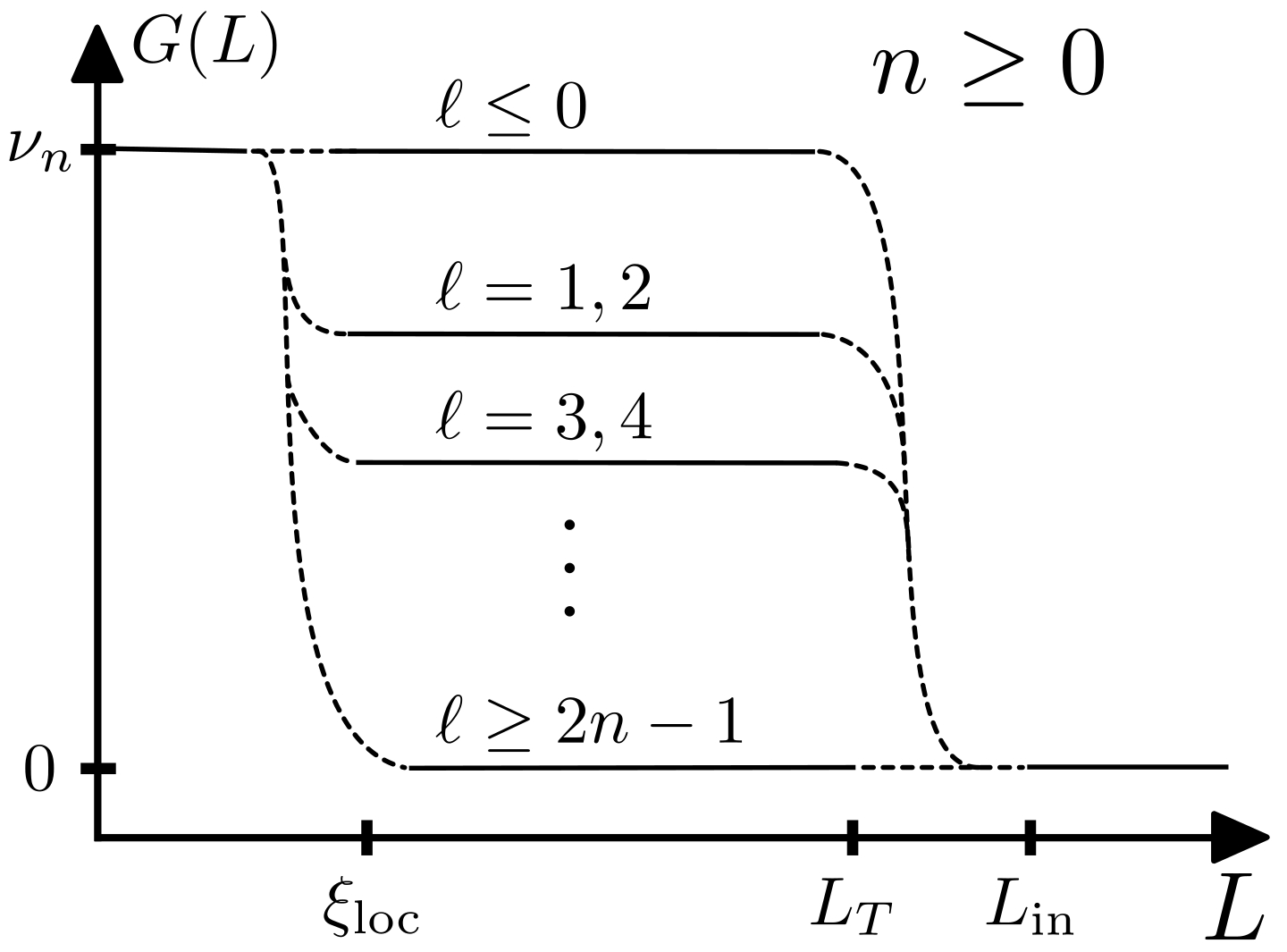}
 \caption{The renormalization group flow of $\pi$-junctions with the $\nu=\frac{2}{5}$ Jain state and paired states with $\ell<0$. Each fixed point corresponds to a specific boundary condition for the neutral modes $\phi_{\sigma,a}$. Single and Double-Andreev refers to the number of neutral modes whose sign changes across the constrictions, which coincides with the number of relevant perturbations. The only stable fixed point is the trivial one, where both neutral modes are continuous.}
 \label{fig.3} 
 \end{figure} 

At the trivial fixed point, the scattering matrix $\hat{\cal S}$ in the basis $(\eta_1,\ldots,\eta_{2n},\gamma_1,\ldots,\gamma_{|\ell|})$ is 
\begin{align}
    \hat{{\cal S}}_\text{triv}=   
    \begin{pmatrix}
   \hat{\mathbb{I}}_{2n}&\\
    &\hat{\mathbb{I}}_{|\ell|}
\end{pmatrix}.
\end{align}
For $\ell \leq 0$, all Majorana co-propagate, and the fixed point is stable. Moreover, any scattering matrix preserving the total number of neutral fermions $n_\psi \equiv \sum_a \psi_{\sigma,a}^\dagger\psi_{\sigma,a} \sim \sum_a \partial_x \varphi_{\sigma,a}$ can be brought into this form by a relative rotation of the Majoranas in the leads that preserves the action of the leads. Generic rotations that change $n_\psi$ introduce non-quadratic terms in the bosonized form of Eq.~\eqref{eq.ch_sigma}. We find that $n_\psi$ breaking perturbations to ${\cal S}_\text{triv}$ are irrelevant under renormalization by the leads while $n_\psi$ preserving ones are marginal; see Appendix~\ref{app.Fixed points}.

For $\ell>0$, the $\eta$ and $\gamma$ Majoranas counter-propagate, and the Majorana mass terms in Eq.~\eqref{eq.L_tun0}, i.e., ${\cal L}_\text{tun}\propto i\gamma_k\eta_l$, localize the maximal number $\text{min}(|\ell|,2n)$ of Majoranas. In this case, the trivial fixed point is unstable. The disorder is a relevant perturbation that drives the system to a maximally localized fixed point with the counter-propagating Majoranas back-scattered. Any $n_\psi$ preserving scattering matrix describing the localized constriction can be brought into the form
\begin{align}
    \hat{{\cal S}}_\text{loc}=   
    \begin{pmatrix}
    \hat{\mathbb{I}}_{2n-\ell}&&\\
    &&\hat{\mathbb{I}}_{\ell}\\
    &\hat{\mathbb{I}}_{\ell}&
\end{pmatrix}\text{  or }
\begin{pmatrix}
    &\hat{\mathbb{I}}_{2n}&\\
    \hat{\mathbb{I}}_{2n}&&\\
    &&\hat{\mathbb{I}}_{\ell-2n}
    \end{pmatrix},
    \label{eqn.s.loc}
\end{align}
for $2n\geq|\ell|$ or $2n<|\ell|$, respectively, without modifying the leads. As before, $n_\psi$ breaking perturbations of the maximally localized fixed points are also irrelevant, and $n_\psi$ preserving ones are marginal.

To identify additional fixed points, we observe that the specific $n_\psi$ breaking transformation $\eta_{2a} \leftrightarrow \eta_{2a+1}$ in Eq.~\eqref{eq.ch_sigma} also yields a quadratic bosonized theory. Such transformations are needed to bring scattering matrices with Andreev processes $\psi_{\sigma,a,\text{in}}\rightarrow \psi^\dagger_{\sigma,a,\text{out}}$ into the canonical form of Eq.~\eqref{eqn.s.loc}. The transformation changes the sign of $h^\sigma_a$ in Eq.~\eqref{eq.ch_sigma}, which modifies various scaling dimensions compared to the $n_\psi$ preserving case. The additional fixed points are thus classified by the number of Andreev processes, and we find a relevant operator for each Andreev process; see Fig.~\ref{fig.3} and Appendix \ref{app.Fixed points}.

We compute the conductance at each fixed point from the current-balance equations. The source S at voltage $V$ supplies a total charge current $I_\text{tot}= \sum_{a=1}^n t_a I^S_a = \nu_n V$, which is equally distributed among the $n$ modes when they are fully equilibrated.

Likewise, we assume that the leads D2 and D3 are equilibrated, and the grounded contact D3 imposes $I^\text{D3}_0\equiv I_0(x_R)=0$. \footnote{For $|\ell|\geq2$, the Majorana modes form complex fermions $\gamma_{2k-1}+i\gamma_{2k}\propto e^{i\phi_k}$ that could carry $U(1)$ current $I_{k}=i\partial_\tau \phi_k/2\pi$ (not to be confused with the thermal current). However, these currents are not protected by charge conservation and dissipate in the presence of disorder. Thus, the impinging neutral $U(1)$ currents on the constriction are zero at $x_R$($x_L$) for $\ell\geq2$($\ell\leq-2$).  If these currents were non-zero $\sum_k I_k=I_{U(1)}$, this would result in induced charge currents $I_\text{D1}=-I_\text{D2}\propto I_{U(1)}$ for source voltage $V=0$.}

It remains to determine the currents $I^{D1}_a\equiv I_a(x_R)$ and $I^{D2}_0=I_0(x_L)$ from the boundary condition imposed by the constriction and encoded in ${\cal S}$. We use the charge-neutral basis for the currents impinging the construction from left and right, which are related to the leads' currents via
\begin{align}\label{eq.basis change}
    (I^\text{L}_\text{ch},I^\text{L}_{\sigma,1},\ldots)=&W^{-1}(I^\text{D2}_0,I^\text{S}_{1},\ldots,I^\text{S}_{n}),\\
    (I^\text{R}_\text{ch},I^{\text{R}}_{\sigma,1},\ldots)=&W^{-1}(I^\text{D3}_0,I^\text{D1}_{1},\ldots,I^\text{D1}_{n}),
\end{align}
with $W$ of Eq.~\eqref{eq.W}. Charge conservation requires $I^\text{L}_\text{ch} = I^\text{R}_\text{ch}$. For $\ell\leq 0$, all $\eta$s propagate in the same direction as $\gamma$s, and all neutral modes $\varphi_{\sigma,a}$ are transmitted. Hence, at the trivial fixed point, the neutral currents are continuous $I^\text{L}_{\sigma,a}=I^\text{R}_{\sigma,a}$. We solve these $n+1$ continuity equations, together with the $n+1$ boundary conditions imposed by the S and D3 leads, and find that the conductance for $\ell\leq 0$ saturates the ballistic conductance value $G_0=\nu_\text{Jain}=\frac{n}{2n+1}$.

For $\ell>0$, the counter-propagating $\gamma$ and $\eta$ Majorana modes localize each other. When at least one of the modes $\eta_{2a-1}$ or $\eta_{2a}$ is localized, the right-moving current $I_{\sigma,a}(x)\propto e^{(x_L-x)/\xi}$ exponentially decays and does not reach the right end of the constriction $I^R_{\sigma,a}\approx 0$. At the maximally localized fixed point, $m(\ell)=\text{min}(\left\lfloor \frac{\ell+1}{2}\right\rfloor,n)$ currents vanish; without loss of generality, we take them as corresponding to $a=1,\ldots,m$. The remaining $n-m(\ell)$
currents are continuous at the constriction. For $\ell<0$, Majorana modes co-propagate, and all the currents are continuous at the trivial fixed point corresponding to $m(\ell)=0$. Together with the charge conservation, we obtain $n+1$ equations
\begin{align}
    I_\text{ch}^\text{L} = I_\text{ch}^\text{R}~,\quad
    I_{\sigma, a \leq m}^{\text{R}}=0~,\quad
    I_{\sigma,a >m }^{\text{L}} = I_{\sigma,a>m}^{\text{R}}~,
    \label{eqn.threebcs}
\end{align}
which we solve using the values for the incoming currents $I^\text{S}_{a=1,\ldots,n}=\frac{V}{2n+1}$ and $I^\text{D3}_0=0$. We find that the coherent charge conductance is
\begin{align}\label{eq.cond}
G_\ell=\frac{n}{2n+1}-\frac{m(\ell)}{2m(\ell)+1},
\end{align}
see Appendix~\ref{app.pos_n} for details. Surprisingly, the conductance at the interface between $\nu=\frac{1}{2}$ states and Jain states with $\nu=\frac{n}{2n+1}$
takes the value of the equilibrated conductance between $n$th and $m(\ell)$th Jain states. Equivalently, the charge current $I^\text{D2}_0$ at the paired edge depends only on the number of Majoranas $\ell$ and not on the Jain-state filling for sufficiently large $n$. 

The analysis of interfaces with hole-conjugate Jain states $n<0$ is performed in Appendix~\ref{app.neg_n}, and the resulting values are listed in Table~\ref{tab.condunctance}. Remarkably, a single electric conductance measurement for suitable $n$ narrows the candidate topological orders down to one non-Abelian and one neighboring Abelian possibility.

\begin{table}[t]
    \centering
        \caption{The coherent charge conductance of different candidate states of $\nu=\frac{12}{5}$ plateau.} 
    \def\arraystretch{1.5}
    \begin{tabular}{c |c c  c  c } \hline\hline
                        & Jain-$\frac{2}{5}$ & $\overline{\text{RR}}$  & BS-MR  & BS-aPf  \\\hline
    $\nu=\frac{8}{3}$   & $\frac{2}{3}$ & $\bf \frac{1}{3}$ & $\frac{2}{3}$ & $\frac{4}{15}$\\ 
    $\nu=\frac{7}{3}$   &  0 & $\frac{1}{3}$ &  $\bf \frac{1}{21}$ & $\frac{1}{3}$ \\\hline\hline
    \end{tabular}
    \label{tab.condunctance_arr}
\end{table}

 The use of the $\pi$-geometry as a litmus test of non-Abelian topological orders extends beyond paired states. In particular, the non-Abelian candidate phases of the $\nu=\frac{12}{5}$ plateau can be identified using experimentally accessible interfaces. Ref.~\onlinecite{Yutushui_daughters_2024} showed that upstream noise could distinguish the anti-Read-Rezayi state from other candidates in the fully charge-equilibrated regime. 

Despite occurring at the same filling factor, these states have different edge theories and, hence, localization properties, which permits $\pi$-geometry to distinguish them. In addition to the requisite charge mode, the anti-Read-Reazy edge contains an upstream Wess-Zumino-Witten $SU(2)_3$ mode, and the Bonderson-Slingerland edge contains a neutral upstream boson mode and $\ell$ Majorana fermions. The non-trivial neutral sectors do not reduce to a free-fermion problem at the constriction. Despite this complication, we find that the maximally localized fixed points analogous to Eq.~\eqref{eqn.s.loc} are again stable, and their coherence conductances can be computed from the Abelian sectors alone; see Appendix~\ref{app.anti_read}. Crucially, the four main candidate phases can be distinguished at the interface between $\nu=\frac{12}{5}$ and the readily available $\nu=\frac{7}{3}$ and $\nu=\frac{8}{3}$ states; see Table~\ref{tab.condunctance_arr}.

{\it Discussion.---}We proposed an experimental protocol for distinguishing pairing channels of the $\nu=\frac{1}{2}$ state based on Table~\ref{tab.condunctance}. This method uniquely identifies the $K=8$ state for which $\ell=0$. More generally, it narrows the possible topological orders down to one non-Abelian state with $|\ell|$ and one Abelian state with $|\ell|+1$ Majoranas. Additionally, the upstream noise measurements proposed in Ref.~\onlinecite{Yutushui_Identifying_2023} single out non-Abelian--Abelian pairs with $|\ell|,|\ell|-1$. Together, these two measurements uniquely identify the topological order.

The appealingly simple form of Eq.~\eqref{eq.cond} prompts several remarks. First, the apparent conversion of neutral Majorana fermions into charge-carrying modes resembles a similar phenomenon occurring in daughter states of paired quantum Hall states, whose filling factor is determined by the parent's $\ell$.~\cite{Yutushui_daughters_2024,Zheltonozhskii_daughters_2024,zhang_hierarchy_2024} Second, Eq.~\eqref{eq.cond} can be understood from the composite fermion picture of these states. In this framework, the paired state is viewed as a superconductor of composite fermions with $\ell$ Majorana modes, which localize with $m(\ell)$ integer modes of the Jain state. Similar localization effects occur at interfaces between the $n$th and $m$th Jain states, yielding the same conductance expression in Eq.~\eqref{eq.cond}. Third, analogous interfaces between 2D topological superconductors and integer quantum Hall states of electrons do not yield a universal conductance \cite{Kurilovich_Disorder_2023} because the Andreev processes are marginal. Consequently, the third equation in \eqref{eqn.threebcs} no longer holds.

Finally, it would be interesting to analyze similar interfaces between two non-Abelian phases where interactions between the neutral sectors result in a qualitatively different edge theory \cite{Lichtman_bulk_anyons_2021}. For example, the counter-propagating neutral sectors of the anti-Pfaffian and anti-Read-Rezayi states can lead to a tricritical Ising theory~\cite{DiFrancesco_CFT_1997,Grosfeld_tricritial_2009} that affects the charge conductance in the $\pi$-geometry.

{\it Acknowledgments.---} MY regards to Alexander D. Mirlin, Jinhong Park, and Christian Spånslätt. This work was supported by grants from the ERC under the European Union’s Horizon 2020 research and innovation programme (Grant Agreements LEGOTOP No. 788715), the DFG (CRC/Transregio 183, EI 519/71), the Israel Science Foundation (ISF) under the Quantum Science and Technology program and ISF grant 2572/21.

\section{End Materials: Green's functions of chiral edge states}\label{app.Green}
In the main text, we assessed the stability of different fixed points based on the scaling dimensions of local perturbations. For the trivial fixed point without localization, the density-density interactions $V_{ab}$ are homogeneous, and the scaling dimensions can be computed by standard methods.~\cite{Moore_Classification_1997} At fixed points with Andreev processes, the boundary conditions $\phi_{\sigma,a}(x_L)=-\phi_{\sigma,a}(x_R)$ can be absorbed into a redefinition of the field on one side of the constriction. This transformation changes the sign of the coupling constants $h_{\sigma,a}\to -h_{\sigma,a}$ for the corresponding field, which makes the density-density interactions $V_{ab}(x)$ position dependent. Similarly, at fixed points with localized modes, the unfolding technique introduced in Appendix~\ref{app.neg_n} results in a continuous neutral sector with inhomogeneous interactions. To evaluate the scaling dimension in such cases, we compute the chiral Green's function; see also Refs.~\onlinecite{Maslov_Landauer_conductance_1995,Rosenow_signatures_2010,Protopopov_transport_2_3_2017,Yutushui_Identifying_2022}. The key step is to impose the proper boundary condition at $x\to\pm \infty$, which are dictated by the topological properties of the chiral edge.

\subsection{Homogeneous case and general properties}

The Lagrangian density of a multi-component chiral Luttinger liquid in imaginary time is 
\begin{align}\label{eq.L_app}
     {\cal L}_{K}  = \frac{1}{4\pi} \partial_x\vect{\phi}(i \hat{K}\partial_\tau-\hat{V}(x)\partial_x)\vect{\phi},
\end{align}
where we adopted a matrix notation for brevity. 
The Matsubara Green's function of Eq.~\eqref{eq.L_app} with constant $V(x)=V$ can be evaluated as 
\begin{align}
     \hat{D}_\omega(x,x_0)\equiv\langle \vect{\phi}(x,\omega)\vect{\phi}^T(x_0,-\omega)\rangle
    =\int \frac{dq }{2\pi} \frac{e^{i q (x-x_0) }}{i\omega q\hat{K} + q^2 \hat{V}}.
\end{align}
 For large $x$, the dominant contributions to the integral arise from small $q$. We neglect the subleading term $q^2 \hat{V}$ to obtain the asymptotic behavior, which is independent of the non-universal matrix $\hat{V}$. The remaining integral is readily evaluated to
\begin{align}
    \hat{D}_\omega(x\to \pm \infty,0) \approx \int \frac{dq }{2\pi} \frac{e^{i q x }}{i\omega q\hat{K} } =\frac{\text{sign($x$)}}{2\omega} \hat{K}^{-1}.
\end{align}
These equations serve as boundary conditions for Green's functions. Crucially, they only depend on the matrix $K$, which is determined by the topological order of the bulk. In an operator framework, it encodes the commutation relations of the quantized fields $[\hat{\phi}_a(x),\hat{\phi}_b(x')]= i \pi \text{sign($x$-$x'$)}\hat{K}^{-1}_{ab}$.

The equation of motion of Eq.~\eqref{eq.L_app} is
\begin{align}\label{eq.EOM}
    \partial_x\left[\omega \hat{K} + \hat{V}(x)\partial_x\right]\vect{\phi}(x) = 0.
\end{align} 
In the regions of constant $\hat{V}$, the solutions are given in terms of the generalized eigenvalues $\lambda_j$ and eigenvectors $\vect{f}_j$ obeying
\begin{align}\label{eq.app_eig}
    \hat{V}\vect{f}_j = \lambda_j \hat{K}\vect{f}_j.
\end{align}
The matrix $\hat{V}$ is positive definite, and the numbers of positive and negative $\lambda_a$ are determined by the signature of $\hat{K}$, i.e., the numbers of left and right moving modes. Then, the solutions to Eq.~\eqref{eq.EOM} are
\begin{align}\label{eq.app_F}
    \vect{\phi}_\omega(x) = \vect{a} + \hat{F}(x)\vect{b},\qquad[\hat{F}(x)]_{ij}=[\vect{f}_j]_i e^{-\frac{\omega x}{\lambda_j}},
\end{align}
where $\vect{a}$ and $\vect{b}$ are vectors of free parameters to be determined by the boundary conditions. 

\subsection{Green's function with inhomogeneous density interactions}\label{app.Green}
The Green's function obeys the differential equation
\begin{align}
    \partial_x\left[ \omega \hat{K} + \hat{V}(x)\partial_x\right]\hat{D}_\omega(x,x_0) = \mathds{1}\delta(x-x_0),
\end{align}
with the boundary conditions derived above, i.e.,
\begin{align}\label{eq.asymptot}
    \hat{D}_\omega(\pm \infty,x_0)=\pm \frac{1}{2\omega} \hat{K}^{-1}.
\end{align}
These equations generalize the Maslov-Stone solution~\cite{Maslov_Landauer_conductance_1995} for the non-chiral field $\varphi$ field; their solution is reproduced by $\hat{K}=2\sigma_x$ in the basis $\vect{\phi}=(\varphi,\theta)$. 

We now assume that $\hat{V}(x) = \Theta(-x)\hat{V}_1 +\Theta(x)\hat{V}_2$ and solve within the $x<0$ and $x>0$ regions separately. The solution for $x_0>0$ is given by
\begin{align}\label{eq.app_G}
    \hat{D}_\omega(x,x_0) =
     \begin{cases}
         -\frac{1}{2\omega} \hat{K}^{-1} + \hat{F}_1(x) \hat{A},\quad &x<0<x_0\\
         \hat{B} + \hat{F}_2(x) \hat{C},\quad &0<x<x_0\\
         \frac{1}{2\omega} \hat{K}^{-1} + \hat{F}_2(x) \hat{E},\quad &0<x_0<x\\
    \end{cases},    
\end{align}
where $\hat{F}_{1(2)}(x)$ is the solution 
Eq.~\eqref{eq.app_F} to Eq.~\eqref{eq.app_eig} with $\hat{V}_{1(2)}$, and $\hat{A},\hat{B},\hat{C}$ and $\hat{E}$ are matrices of parameters. The boundary conditions Eq.~\eqref{eq.asymptot} require that $A_{ij}=0$ for $\lambda_j>0$ and $E_{ij}=0$ for $\lambda_j<0$. The remaining elements of $\hat{A}$, $\hat{E}$, and the coefficient matrices $\hat{B}$ and $\hat{C}$ are determined from (i) the continuity of $\hat{D}_\omega(x,x_0)$ at $x=0$ and $x_0$; (ii) the continuity of $\hat{V}(x)\partial_x \hat{D}_\omega(x,x_0)$ at $x=0$; and (iii)
the unit discontinuity of $\hat{V}(x)\partial_x \hat{D}_\omega(x,x_0)$ at $x=x_0$. Explicitly, the conditions are 
\begin{align}\label{eq.app_BC}
    \text{(i)}&\qquad \hat{D}_\omega(x,x_0)|_{-0^+}^{0^+}=\hat{D}_\omega(x,x_0)|_{x_0-0^+}^{x_0+0^+}=0,\\
    \text{(ii)}&\qquad \hat{V}_1\partial_x\hat{D}_\omega(0^-,x_0)=\hat{V}_2\partial_x\hat{D}_\omega(0^+,x_0),\\
    \text{(iii)}& \qquad \hat{V}_2\partial_x \hat{D}_\omega(x,x_0)|_{x=x_0-0^+}^{x=x_0+0^+}=\mathds{1}.
\end{align}
The scaling dimension of $e^{i\vect{m}\vect{\phi}}$ can now be computed as $\Delta_m= \vect{m}\hat{\Delta}\vect{m}$, where 
\begin{align}\label{eq.app_Delta}
    \hat{\Delta} = -\lim\limits_{\omega\to 0}\omega \hat{D}_\omega(x,x_0)
\end{align}
is independent of $x$ and $x_0$.

\appendix

\section{Fixed points of $\pi$-geometry with disorder.}\label{app.Fixed points}
We now analyze the $\pi$-geometry in the presence of random tunneling between the Jain and paired-state edges in the constriction $x\in {\cal R}_\text{Constriction}\equiv [x_L,x_R]$. We first derive the transformations that absorb disorder and identify fixed points. In the main text, such disorder configurations are parametrized by scattering matrices for the Majorana fermions. 

In the leads, $x\in {\cal R}_\text{Leads}\equiv\{x|x<x_L\lor x>x_R\}$, the edge modes of Jain states at $\nu_n=\frac{n}{2n+1}$ and paired states are spatially separated and decoupled. Consequently, the total Lagrangian density of the leads is given by
\begin{align}\label{eq.app_lead}
    {\cal L}_\text{Leads}=\frac{1}{4\pi}\partial_x \vect{\phi}\left( i \hat{K}\partial_\tau + \hat{V}_\text{Leads}\partial_x\right)\vect{\phi}+{\cal L}_\ell,
\end{align}
with a block-diagonal interaction matrix $V_\text{leads} =\text{diag}(v_0,V_\text{Jain})$ and  ${\cal L}_\ell$ specified in Eq.~\eqref{eq.Lgamma} of the main test. In the charge-neutral basis of Eq.~\eqref{eq.W} in the main text, the charge modes and neutral modes interact via $h_{\sigma}\partial_x\varphi_\text{ch} \partial_x \varphi_{\sigma,a}$ with the specific coupling strength $h_\sigma=-\frac{4n+2}{4n+1}(v_\text{ch} + v_\sigma)$. The $n$-dependent factor is the inverse filling factor difference between the Jain and paired states, i.e., $\delta \nu^{-1}=(4n+2)$. 

For $x\in {\cal R}_\text{Constriction}$, we adopt a fermionic formulation and introduce Majorana modes $\eta_k$ via $e^{i\varphi_{\sigma,a}}=\eta_{2a-1}+i\eta_{2a}$. We collect all $\eta$ Majoranas and the $\gamma$ Majoranas of the paired state into the vector $\vect{\Psi}=(\eta_1,\ldots\eta_{2n},\gamma_1,\ldots,\gamma_{|\ell|})$. In this basis, the Lagrangian Eq.~\eqref{eq.app_lead} is given by
\begin{align}
         {\cal L}^\text{ch}_\text{Leads}&=\frac{1}{4\pi} \;
        \partial_x \varphi_\text{ch}\left( i\delta \nu^{-1} \partial_\tau + v_c\partial_x\right)\varphi_\text{ch}\\
          {\cal L}^{\Psi}_\text{Leads}&= \vect{\Psi}^T\left(\mathbb{I}_n\partial_\tau - i \hat{V}_\Psi \partial_x\right)\vect{\Psi}, \\
          {\cal L}^\text{int}_\text{Leads}&=h_{\sigma} \; \partial\varphi_\text{ch}\vect{\Psi}^T\hat{\Sigma}_y\vect{\Psi},
\end{align}
 with the matrix $\hat{\Sigma}_y=\text{diag}(\sigma_y,\ldots,\sigma_y,0,\ldots ,0)$ such that $\frac{1}{2\pi}\sum_a \partial_x \varphi_{\sigma,a}=\vect{\Psi}^T\hat{\Sigma}_y\vect{\Psi}$.

\subsection{Co-propagating $\eta$ and $\gamma$ Majoranas.}
We begin with $n \ell<0$, where all Majoranas co-propagate. Here, we do not discriminate between particle-like ($n>0$) and hole-like ($n<0$) Jain states since both cases map to a theory with a single charge mode $\varphi_\text{ch}$ and $2|n|+|\ell|$ co-propagating Majorana fermions moving in the opposite to the direction of charge. 

The Lagrangian of the leads is significantly constrained by the spatial separation between the Jain and paired-state edges, which implies the absence of interactions or tunneling between them. By contrast, any local and charge-conserving terms are allowed in the constriction. We organize the general Lagrangian of the constriction into four contributions that anticipate the fixed point, i.e., ${\cal L}_\text{Cons} = {\cal L}^\text{ch}_\text{Cons}+ {\cal L}^{\Psi}_\text{Cons}+  {\cal L}^\text{dis}_\text{Cons}  + {\cal L}^\text{pert}_\text{Cons}$ with
\begin{align}
         {\cal L}^\text{ch}_\text{Cons}&=\frac{1}{4\pi} \;
        \partial_x \varphi_\text{ch}\left( i\delta \nu^{-1} \partial_\tau + \tilde{v}_c\partial_x\right)\varphi_\text{ch},\\
          {\cal L}^{\Psi}_\text{Cons}&=\vect{\Psi}^T\left(\partial_\tau + i  v_n \partial_x\right)\vect{\Psi},\\
    {\cal L}^\text{dis}_\text{Cons}&= \vect{\Psi}^T\hat{\cal G}(x)\vect{\Psi},\\
     {\cal L}^\text{pert}_\text{Cons}&=
  \vect{\Psi}^T\delta \hat{V}\partial_x\vect{\Psi} + \; \partial\varphi_\text{ch}\vect{\Psi}^T\hat{M}_1\vect{\Psi}\;\nonumber\\
  &\quad+  \vect{\Psi}^T\hat{M}_2\vect{\Psi}\;\vect{\Psi}^T\hat{M}_3\vect{\Psi}.
\end{align}
The first two terms describe ballistic propagation of charge modes $\varphi_\text{ch}$ and neutral fermions $\Psi$ with the velocities $\tilde v_c$ and $v_n$, respectively. The third term, ${\cal L}^\text{dis}_\text{Cons}$, represents the random tunneling between electrons on the Jain and paired states edges parametrized by a skew-symmetric matrix  $\hat{\cal G}(x)$. The final term collects the most relevant perturbations. First, a velocity anisotropy of the Majorana modes, parameterized by an anti-symmetric matrix $\delta \hat{V}$. Second, charge-neutral and neutral-neutral interactions are parameterized by three symmetric matrices $\hat{M}_{1,2,3}$.

The random tunneling terms ${\cal L}^\text{dis}_\text{Cons}$ can be absorbed by a disorder-specific transformation $\tilde{\vect\Psi}(x) = \hat{\cal U}(x)\vect{\Psi}(x)$ with
\begin{align}\label{eq.app_trans}
     \hat{\cal U}(x) = \begin{cases}\hat{\cal U}_L&:\;x<x_L,\\
         \hat{\cal U}_L \;\hat{\cal U}_{\hat{\cal G}}(x) &:\;x\in{\cal R}_C,\\
         \hat{\cal U}_R&:\;x>x_R,
     \end{cases}.
\end{align}
where $\hat{\cal U}_{\hat{\cal G}}(x)=T_x\exp\left[\frac{i}{v_n}\int_{x_L}^{x} \hat{\cal G}(y) dy\right]$, and $T_x$ denotes path ordering. Here, $\hat{\cal U}_{L/R}\in O(2n+\ell)$ are constant transformations that we can choose such that $\hat{\cal U}(x)$ is smooth. This transformation absorbs ${\cal L}^\text{dis}_\text{Cons}$ into ${\cal L}^{\Psi}_\text{Cons}$. In the new basis, the couplings in ${\cal L}^\text{pert}_\text{Cons}$ become random variables. If the remaining theory is quadratic {\it everywhere}, the Lagrangian ${\cal L}^\text{pert}_\text{Cons}$ is irrelevant, and the theory flows to decoupled charge and neutral sectors with equal velocities of Majorana modes. 

The transformation $\hat{\cal U}(x)$ also changes the Lagrangian in the leads since the terms 
\begin{align}\label{eq.app_term}
    \partial\varphi_\text{ch}\vect{\Psi}^T\delta\hat{\Sigma}_y\vect{\Psi} = 2i \partial\varphi_\text{ch} \sum_{k=1}^{n} \eta_{2k-1}\eta_{2k}
\end{align}
are not invariant under $\vect{\Psi}\to \hat{\cal U}_{L/R}\vect{\Psi}$ for arbitrary $\hat{\cal U}_{L/R}$. For instance, if $\hat{\cal U}_{L/R}$ mixes $\eta_1$ and $\eta_3$ Majoranas, the term $\partial\varphi_\text{ch}\cos\sqrt{2}(\varphi_{\sigma,1}-\varphi_{\sigma,2})$ is generated. However, if the disorder configuration $\hat{\cal G}_0(x)$ is such that $\hat{\cal U}_{L}$ and $\hat{\cal U}_{R}$ preserve the quadratic form of the lead Lagrangian, the theory is at a fixed point perturbed by ${\cal L}^\text{pert}_\text{Cons}$.

\paragraph{Fixed points.} We first identify transformations that correspond to fixed points and then study the relevance of the perturbations ${\cal L}^\text{pert}_\text{Cons}$ and $\vect{\Psi}^T\delta\hat{\cal G}(x)\vect{\Psi}$. We identify two classes of such transformation: Transformations that preserve $n_\psi=\vect{\Psi}^T\hat{\Sigma}_y\vect{\Psi}$ and transformations that contain Andreev processes. 

The $n_\psi$ preserving transformations are given by any combination of (i) any rotation between $\gamma_k$ Majoranas of the paired state, (ii) any rotation within the subspace of $(\eta_{2k-1},\eta_{2k})$, and (iii) simultaneous rotation of $(\eta_{2k_1-1},\eta_{2k_2-1})$ and $(\eta_{2k_1},\eta_{2k_2})$.  Such transformations do not modify the Lagrangian and are merely a basis change. The disorder perturbations that preserve $n_\Psi$ are thus exactly marginal.

The Andreev transformations preserve the quadratic form of the action but modify the density-density interactions. The transformations that exchange two Majoranas $\eta_{2a-1}\leftrightarrow\eta_{2a}$ correspond to $\partial_x\varphi_{\sigma,a} \to -\partial_x\varphi_{\sigma,a}$. These transformations change the sign of the corresponding density-density interaction parameter $h_{\sigma,a}\to-h_{\sigma,a}$ on one side of the constriction.

\paragraph{Perturbations.} Firstly, we consider disorder perturbations $\hat{\cal G}_0(x)\to \hat{\cal G}_0(x)+ \delta\hat{\cal G}(x)$. Suppose the disorder $\hat{\cal G}_0(x)$ is such that it is fully absorbed by $\hat{\cal U}_0(x)$, which corresponds to a fixed point. Then, a similar transformation $\hat{\cal U}(x)$ can collect the disorder perturbation $\delta\hat{\cal G}(x)$ to a single point,~\cite{Protopopov_transport_2_3_2017,Yutushui_Identifying_2022} which for concreteness we take as $x_R$. Explicitly, such transformation is given by
\begin{align}\label{eq.app_trans}
     \hat{\cal U}(x) = \begin{cases}\hat{\cal U}_{L0}&:\;x<x_L.\\
         \hat{\cal U}_{L0}\; \hat{\cal U}_{\hat{\cal G}}(x)&:\;x\in{\cal R}_C\\
         \hat{\cal U}_{L0}\hat{\cal U}_{\hat{\cal G}}(x_R)T_x e^{\frac{i\delta \hat{\cal G}}{v_n}\frac{(x_R-x)}{\epsilon}}&:\;x\in[x_R,x_R+\epsilon]\\
         \hat{\cal U}_{R0}&:\;x>x_R+\epsilon,
     \end{cases},
\end{align} 
where $\epsilon>0$ is infinitesimal, and $\delta\hat{\cal G}$ is such that
\begin{align}
    T_x\exp\left[\frac{i\delta\hat{\cal G}}{v_n}\right] =  \hat{\cal U}_{\hat{\cal G}}^T(x_R)\; \hat{\cal U}_{\hat{\cal G}_0}(x_R).
\end{align}
Notice that $\hat{\cal U}_{L/R,0}$ in $\hat{\cal U}(x)$ are the same as in $\hat{\cal U}_0(x)$. However, such transformations introduce a point-like tunneling ${\cal L}_\text{Dis}^\text{pert}=\vect{\Psi}^T\delta \hat{\cal G}\vect{\Psi}$. The stability of the fixed point is determined by the relevance of such perturbations. We calculate their scaling dimensions using Green's function, as discussed in the End Material.

We find that the perturbation ${\cal L}_\text{dis}^\text{pert}$ is irrelevant at the trivial fixed point. Likewise, the second and third terms in ${\cal L}^\text{pert}_\text{Cons}$ are irrelevant. Regarding the first term in ${\cal L}^\text{pert}_\text{Cons}$, the transformation $\hat{\cal U}(x)$ replaces $\delta \hat V$ by a random disorder-dependent matrix, rendering it irrelevant. In contrast,  we find that at the fixed points that include Andreev processes, the perturbation ${\cal L}_\text{Dis}^\text{pert}$ is relevant, which implies that they are unstable.

\subsection{Counter-propagating $\eta$ and $\gamma$ Majoranas.}
\begin{figure}
    \centering
    \includegraphics[width=0.95\linewidth]{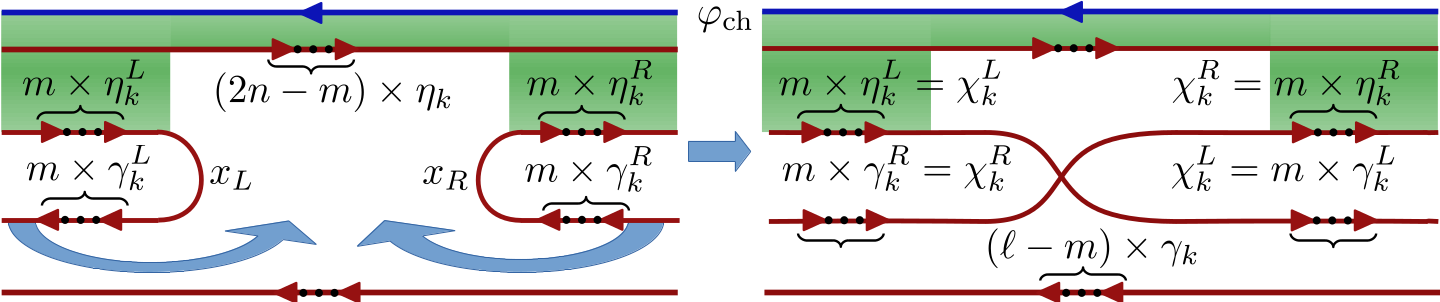}
    \caption{We use an `unfolding' transformation to map a constriction with localized (back-reflected) fermion modes onto a problem where all fermions are transmitted. Each of the resulting fermions $\chi$, defined in Eq.~\eqref{eq.app_unfold}, interacts with the charge mode only on one side of the constriction as indicated by the green regions. }
    \label{fig.unwarp}
\end{figure}

For $n \ell>0$, $\eta$ and $\gamma$ Majoranas counter-propagate and localize in pairs. To describe this situation on the same footing as above, we perform an `unfolding' transformation shown in Fig.~\ref{fig.unwarp}.

Any localizing pair $\gamma$ and $\eta$ obeys the back-scattering boundary conditions $\gamma_k(x_{R/L})=\eta_k(x_{R/L})$. We now define two new sets of chiral Majorana modes 
\begin{align}\label{eq.app_unfold}
    \chi^L_k=\theta(x_L-x)\eta_k(x) + \theta(x-x_L)\gamma_k(x_L-x),\\
    \chi^R_k=\theta(x-x_R)\eta_k(x) + \theta(x_R-x)\gamma_k(x_R-x).
\end{align}

After this transformation, $\chi$ and $\eta$ Majoranas move in the same direction. Recall that, in the leads, the $\eta$ Majoranas couple to the charge mode $\varphi_\text{ch}$ with a specific coupling strength, which corresponds to decoupled paired and Jain state edges. By contrast, the $\gamma$ Majoranas are decoupled from $\eta$ Majoranas and charge mode. Consequently, the unfolded Lagrangian is local but exhibits different couplings to the charge mode on either side of the constriction; see Fig.~\ref{fig.unwarp}. Using this transformation and Green's function calculation explained in the End Materials, we find that the maximally localized fixed point with no Andreev transmission for $\eta$ Majoranas is the only stable fixed point.  

\section{Example calculations of fixed-point stability}
We now exemplify calculations of the scaling dimensions of the disorder tunneling perturbations to assess the stability of the fixed points for the interfaces of the paired and the Jain states at $\nu=\frac{1}{3}$ and $\nu=\frac{2}{5}$. The general $n$ and $\ell$ cases can be verified along the same lines. A similar analysis for the case of the $\nu=\frac{2}{3}$ state was carried out in Ref.~\onlinecite{Protopopov_transport_2_3_2017}. For the $\pi$-geometry geometry with half-filled and integer interfaces, the calculation can be found in Ref.~\onlinecite{Yutushui_Identifying_2022}. 

\subsection{PH-Pfaffian - $\nu=\frac{1}{3}$ interface}
We start with the case of PH-Pfaffian and the $\nu=\frac{1}{3}$ Laughlin state. The edge theory is described by ${\cal L}={\cal L}_\text{lead}+{\cal L}_\text{cons}+{\cal L}_\gamma +{\cal L}^\text{dis}_\text{Cons}$ with
\begin{align}
    {\cal L}_\text{Lead}&= \frac{1}{4\pi}\partial_x\vect{\phi} \left(i\hat{K}\partial_\tau + \hat{V}_\text{Lead}\partial_x\right)\vect{\phi},\\
    {\cal L}_\text{Cons} &= \frac{1}{4\pi}\partial_x\vect{\phi} \left(i\hat{K}\partial_\tau + \hat{V}_\text{Cons}\partial_x\right)\vect{\phi},\\
    {\cal L}_\gamma &= i\gamma (i\partial_\tau - v_\gamma\partial_x)\gamma,\\
    {\cal L}^\text{dis}_\text{Cons} &=  g(x) \gamma e^{i\vect{m}\vect{\phi}} +\text{H.c.},
\end{align}
where $\hat{K}=\text{diag}(-3,2)$, charge vector $\vect{t}=(1,1)$. The final term with $\vect{m}=(3,2)$ represents the tunneling of charge one from the $\nu=\frac{1}{3}$ edge to the $\nu=\frac{1}{2}$ edge. Since the edge modes $\phi_0$ and $\phi_1$ are spatially separated in the leads $x\in {\cal R}_L$, the density-density interaction $h_\text{Lead}\partial_x\phi_1\partial_x\phi_2$ vanishes, and the velocity matrix is diagonal in this basis
\begin{align}
    \hat{V}_\text{Lead}\equiv\begin{pmatrix}
        3v_1 & h_\text{Lead}\\
        h_\text{Lead} & 2v_2\\
    \end{pmatrix}=\begin{pmatrix}
        3v_1 & 0\\
        0 & 2v_2\\
    \end{pmatrix}.
\end{align}

In the constriction, $x\in {\cal R}_C$, the charge and neutral modes $(\varphi_\text{ch}, \phi_\sigma) = W(\phi_1,\phi_2)$ with
\begin{align}
    W=\begin{pmatrix}
        -2 & 1\\
        3 & -1\\
    \end{pmatrix},
\end{align}
are decoupled. Thus, the velocity matrix obeys $\hat{V}_\text{Cons}=W^{T}\text{diag}(6 v_\text{ch},v_\sigma)W$.  

We now study the relevance of the disorder perturbation $\delta g(x)$, which we collect into a single point $x_0$. Then, this perturbation amounts to 
\begin{align}
    {\cal L}^\text{res}_\text{tun} = \delta g \gamma(x_0) e^{i\phi_\sigma(x_0)}.
\end{align} 
Since $\gamma$ is decoupled, its scaling dimension is $\Delta_\gamma = \frac{1}{2}$. Thus, the fixed point is stable if the scaling dimension is $\Delta_\sigma=[e^{i\phi_\sigma}]>\frac{1}{2}$ and unstable if $\Delta_\sigma < \frac{1}{2}$.

The analyses can be carried out with three regions, $x<x_L$, $x\in {\cal R}_C$, and $x>x_R$. However, we can simplify the calculation by neglecting the finite region ${\cal R}_C$ in the zero-frequency limit or, equivalently, the long-distance limit. Hence, we can replace the constriction with a point-like scatterer at $x_0=0$ and a set of boundary conditions of $\phi_{\sigma,a}(0^-)=\pm\phi_{\sigma,a}(0^+)$ corresponding to the trivial or Andreev fixed points. Then, the problem is equivalent to the one considered in the End Materials with $\hat{V}_1$ for $x<0$ and $\hat{V}_2$ for $x>0$.

{\bf Trivial fixed point.} At the trivial fixed point, we solve the Green's function problem Eqs.~\eqref{eq.app_G} with boundary conditions Eq.~\eqref{eq.app_BC} with $\hat{V}_1=\hat{V}_2=\hat{V}_\text{Lead}$. Then, the scaling dimension of neutral fermion $e^{i\phi_\sigma}=e^{i\vect{m}\vect{\phi}}$ is immediately inferred from Eq.~\eqref{eq.app_Delta} with 
\begin{align}
    \hat{\Delta} =\frac{1}{2} \begin{pmatrix}
        \frac{1}{3} & 0\\
        0 & \frac{1}{2}\\
    \end{pmatrix}
\end{align}
to be $\Delta_\sigma =(3,2)^T \hat{\Delta}(3,2)= \frac{5}{2}$. This result can be easily understood since it corresponds to decoupled $\nu=\frac{1}{3}$  and $\nu=\frac{1}{2}$ edges. Thus, $\Delta$ is just a sum of the scaling dimension of three Laughlin quasiparticles, $\frac{3}{2}$, and of two semions, $1$. The scaling dimension of the tunneling term $\gamma e^{i\vect{m}\vect{\phi}}$ at the trivial fixed point is $\Delta = \frac{5}{2} +\frac{1}{2}$, showing that $\delta g$ is irrelevant and the trivial fixed point is stable.

{\bf Andreev transmission fixed point.} At the Andreev transmission fixed point, $\phi_\sigma(x)$ changes sign at $x=0$. Thus, we define a continuous field $\tilde{\phi}_\sigma(x) = \text{sgn}(x)\phi_\sigma(x)$. For $x<0$, the new fields are $(\tilde{\phi}_1,\tilde{\phi}_2)=W(\varphi_\text{ch},-\phi_\sigma) \equiv A (\phi_1,\phi_2)$ where 
\begin{align}
    A = W\begin{pmatrix}
        1 & 0\\
        0 & -1\\
    \end{pmatrix} W^{-1}=\begin{pmatrix}
        -5 & -4\\
        6 & 5\\
    \end{pmatrix}.
\end{align}
The new interaction matrix is $\tilde{V}_1 = A^T V_1 A$. We now compute the Green's function $\hat{\tilde{D}}_\omega(x,x_0)$ in the $\tilde{\vect{\phi}}$ basis as in the End materials with $\tilde{V}_1 = A^T \hat{V}_\text{Lead} A$ and $\tilde{V}_2 = \hat{V}_\text{Lead}$. The Green's function of the original fields $(\phi_1,\phi_2)$ reads $\hat{D}_\omega(x,x_0) = A \hat{\tilde{D}}_\omega(x,x_0)$ for $x<0$ and $\hat{D}_\omega(x,x_0) = \hat{\tilde{D}}_\omega(x,x_0)$ for $x>0$ (note that $A^2=\mathbb{I}$). In particular, the scaling dimensions are inferred from Green's function on the same side of the constriction at $x=0$, i.e., for $x>0$ and $x_0>0$, and the matrix of scaling dimensions is 
\begin{align}
    \hat{\Delta}=\frac{1}{2} \begin{pmatrix}
        \frac{1}{3} & -\frac{4}{5}\\
        0 & \frac{1}{2}\\
    \end{pmatrix}.
\end{align}
In particular, the scaling dimension of the neutral fermion is $\Delta_\sigma =\frac{1}{10}$ at the twisted fixed point. Consequently, $\delta g$ is a relevant perturbation, which implies that the fixed point is unstable, and the system flows to the trivial fixed point with the ballistic conductance value. 

\subsection{PH-Pfaffian - $\nu=\frac{2}{5}$ interface}
In this case, the $K$-matrix can be transformed to the charge-neutral basis $K'=W^TKW=\text{diag}(-10,1,1)$ via
\begin{align}
    K= \begin{pmatrix}
        3 & 2 & 0\\
        2 & 3 & 0\\
        0 & 0 & -2
    \end{pmatrix},\quad
    W=\begin{pmatrix}
        1 & 1 & 1\\
        3 & 2 & 2\\
        2 & 3 & 2
    \end{pmatrix}.
\end{align}
The analysis is very similar to the previous section, with the main change that now there are three fixed points: 
\begin{itemize}
    \item[i]  Trivial fixed point, with both neutral modes continuous.
    \item[ii] Single Andreev transmission fixed point, with one neutral mode changing sign $\varphi_{\sigma,1}(0^-)=-\varphi_{\sigma,1}(0^+)$ and $\varphi_{\sigma,2}(0^-)=\varphi_{\sigma,2}(0^+)$.
    \item[iii] Double-Andreev transmission fixed point, with both neutral modes changing sign $\varphi_{\sigma,1}(0^-)=-\varphi_{\sigma,1}(0^+)$ and $\varphi_{\sigma,2}(0^-)=-\varphi_{\sigma,2}(0^+)$.
\end{itemize}
Each fixed point can be solved with Eq.~\eqref{eq.app_G}, and $\tilde{V}_1 = A^T \hat{V}_\text{Lead} A$  and $\tilde{V}_2 = \hat{V}_\text{Lead}$ with $A=WTW^{-1}$ encoding the type of fixed point, $T_\text{trivial}=\text{diag}(1,1,1)$, $T_\text{single}=\text{diag}(1,1,-1)$, and $T_\text{double}=\text{diag}(1,-1,-1)$. The resulting scaling dimension matrices are 
\begin{align}
        \hat{\Delta} = \frac{1}{2}\begin{pmatrix}
        K^{-1}_\text{Jain} & \vect{d}\\
        0 & K^{-1}_\text{semion}\\
    \end{pmatrix},
\end{align}
with $\vect{d}_\text{trivial}=(0,0)$,
$\vect{d}_\text{single}=(0,-\frac{1}{5})$ and $\vect{d}_\text{double}=(-\frac{1}{9},-\frac{1}{9})$. It follows that only the trivial fixed point is stable, while single and double Andreev transmission fixed points have one and two relevant operators, respectively.

\section{Current calculation of coherent conductances}
In this Appendix, we compute the coherent conductances for interfaces of paired states and Jain states with $n<0$, and prove the general formula Eq.~\eqref{eq.cond} in the main text. Finally, we compute the coherent conductance for the interfaces necessary for the identification of the $\nu=\frac{12}{5}$ plateau.

\subsection{Conductance of paired state and Jain states with $n>0$}\label{app.pos_n}
We solve equation Eq.~\eqref{eqn.threebcs} in the main text for arbitrary $n$ and $m$. Firstly, we rewrite it in a matrix from 
\begin{align}
\begin{pmatrix}
    I^\text{R}_\text{ch}\\
    \vect{I}^\text{R}_{\sigma}
    \end{pmatrix}
    =    \hat{\cal P}_{m}
    \begin{pmatrix}
    I^\text{L}_\text{ch}\\
    \vect{I}^\text{L}_{\sigma}
    \end{pmatrix},
\end{align}
where $\vect{I}^\text{R/L}_{\sigma}$ are $n$ dimensional neutral currents, and $\hat{\cal P}_{m}$ is an $n+1$-dimensional diagonal projection matrix 
\begin{align}
    \hat{\cal P}_{m}=\text{diag}(1,\underbrace{0,\ldots,0}_{m},\underbrace{1,\ldots,1}_{n-m}).
\end{align}
The first entry and the last block correspond to the conservation of charge and of $n-m$ neutral currents, i.e., the first and third parts of Eq.~\eqref{eqn.threebcs} in the main text. The middle block reflects $m$ vanishing currents at the right end of the constriction. 

We recast this equation into the original basis of currents at the Jain and paired edges using Eq.~\eqref{eq.basis change} with Eq.~\eqref{eq.W} in the main text. We obtain
\begin{align}
\begin{pmatrix}
    I^\text{D3}_\text{0}\\
    \vect{I}^\text{D1}
    \end{pmatrix}
    =    W\hat{\cal P}_{m}W^{-1}
    \begin{pmatrix}
    I^\text{D2}_\text{0}\\
    \vect{I}^\text{S}
    \end{pmatrix}
    \equiv\begin{pmatrix}
    T_{00}&\vect{T_{01}}^T\\
    \vect{T_{10}}&\hat{T}_{11}
    \end{pmatrix}
    \begin{pmatrix}
    I^\text{D2}_\text{0}\\
    \vect{I}^\text{S}
    \end{pmatrix},
\end{align}
where $\vect{I}^\text{S}$ and $\vect{I}^\text{D1}$ are $n$ currents on the Jain edge, and $T_{00}$, $\vect{T_{01}}^T,\vect{T_{10}},\hat{T}_{11}$ are $1\times1$, $1\times n$, $n\times 1$ and $n\times n$ matrices, respectively. The matrix product $W\hat{\cal P}_{m}W^{-1}$ for arbitrary $n$ and $m$ is given by 
\begin{align*}
 W\hat{\cal P}_{m}W^{-1}=&
 \begin{pmatrix}
    2m+1 & \cdots & 2m+1 & & 2m & \cdots & 2m \\
    -2 & \cdots & -2 & & -2 & \cdots & -2 \\
    \vdots & & \vdots & & \vdots & & \vdots \\
    -2 & \cdots & -2 & & -2 & \cdots & -2 \\
    0 & \cdots & 0 & & 1 &\cdots & 0\\
    \vdots & & \vdots & & \vdots & \ddots &\vdots \\
    0 & \cdots & 0 & & 0 &\cdots & 1
\end{pmatrix}.\\
&\raisebox{1.2cm}{$\hspace{0.7cm}\underbrace{\hphantom{\hspace{2.3cm}}}_{\text{$m+1$ entries}}
\hspace{0.8cm}
\underbrace{\hphantom{\hspace{1.7cm}}}_{\text{$n-m$ entries}}$}
\end{align*}

We assume that the source and the drain D3 are at voltages $V_S$ and $V_\text{D3}$, respectively. Since the currents entering the constriction from S are equilibrated, $\vect{I}_S=\vect{t} i_s$, where $t_a=1$ is a $n$-dimensional charge vector of the Jain state, and $i_\text{S}=\frac{V_\text{S}}{2n+1}$ is the current per mode. Multiplying the lower set of equations for $\vect{I}^\text{D1}$ with $\vect{t}^T$, we obtain 
\begin{align}
    I^\text{D3}_\text{0} &= T_{00}I^\text{D2}_\text{0} + \vect{T_{01}}^T\vect{t}\;i_\text{S},\\
    I^\text{D1}_\text{tot} &= \vect{t}^T\vect{T_{10}} I^\text{D2}_\text{0} + \vect{t}^T\hat{T}_{11} \vect{t}\;i_\text{S},
\end{align}
where $ I^\text{D1}_\text{tot}=\vect{t}^T \vect{I}^\text{D1}$ is the total current entering the drain D1. Solving these equations for the outgoing currents $I^\text{D1}_\text{tot}$ and $I^\text{D2}_0$ as a function of $V_\text{S}$ and $V_\text{D3}$, we find
\begin{align}
\label{eqn.sup.c1}
    I^\text{D1}_\text{tot} = &\left(\frac{n}{2n+1}-\frac{m}{2m+1}\right)V_\text{S}
    + \;\frac{m}{2m+1}V_\text{D3},\\
    \label{eqn.sup.c2}
    -I^\text{D2}_0 = &\frac{m}{2m+1}V_\text{S} + \left(\frac{1}{2}-\frac{m}{2m+1}\right)V_\text{D3},
\end{align}
where we use the convention that positive current flows to the right. The conductance from the main text is defined as $G\equiv \frac{I^\text{D1}_\text{tot}}{V_\text{S}}$ at $V_\text{D3}=0$. Remarkably, the conductance between the source S and the drain D2 depends only on the number of neutral modes $\ell$ through $m(\ell)$ and is independent of the incoming current $I^\text{S}_\text{tot}=\nu_n V_\text{s}$ for sufficiently large $n$.

\subsection{Conductance of paired state and Jain states with $n<0$}\label{app.neg_n}
We now compute the conductance for the $n<0$ case. In contrast to the $n>0$ case, the Jain state now occurs at higher filling $\nu_{n}>\nu_\text{Paired}$. Consequently, in the fully equilibrated regime, the conductance is $G_\text{eq}=\nu_{-|n|}-\nu_\text{Paired}$. 

The $K$-matrix of the Jain state is  $K^\text{Jain}_n=2\vect{t}\vect{t}^T - \mathbb{I}_{|n|}$ and has one downstream and $|n|-1$ upstream modes. Since the Jain state for $n<0$ is non-chiral, the conditions that Jain modes are at local equilibrium at the right end of the constriction $x=x_R$ imposes the condition $t_a\mu^\text{D1}_b=t_b\mu^\text{D1}_a$ on the chemical potential of the $a$th mode defined in terms of currents as $\mu^\text{D1}_a = \sum_b K_{ab} I^\text{D1}_b$.~\cite{Yutushui_Localization_2024} The solutions to these conditions, $I^\text{D1}_a=I^\text{D1}_b$, provide $|n|-1$ linearly independent boundary conditions. Together with the total charge current $I^\text{S}_\text{tot}$, there are $|n|$ equations that replace the boundary conditions $I^\text{S}_a =\frac{1}{2n+1} V$ for the positive $n$ case. 

We now proceed similarly to the main text and define a charge-neutral basis  $I_\text{ch}^\text{R/L}$ and $I_{\sigma,a}^{\text{R/L}}$ by Eq.~\eqref{eq.basis change} with
\begin{align}
      W =   
    \begin{pmatrix}
     1 &  \begin{matrix} 1& \cdots \end{matrix} \\
    \begin{matrix} 2\\ \vdots \end{matrix} &  \vcenter{\hbox{\scalebox{1.3}{$K^\text{Jain}_n$}}}
    \end{pmatrix}^{-1}.
\end{align}

If the pairing channel of the half-filled state is negative $\ell<0$, $|\ell|$ upstream Majoranas modes move in the opposite direction to the $|n|$ neutral mode $\varphi_{\sigma,a}$. After localization, $m(\ell)=\text{min}(\left\lfloor\frac{|\ell|+1} {2}\right\rfloor,|n|)$ currents vanish exponentially $I_{\sigma,a\leq m}(x_L)$, while others are continuous at the stable fixed point.
To obtain the conductance, we solve $|n|+1$ equations 
\begin{align}
    I_\text{ch}^\text{L} = I_\text{ch}^\text{R}~,\qquad
    I_{\sigma, a \leq m}^{\text{L}}=0~,\qquad
    I_{\sigma,a >m }^{\text{L}} = I_{\sigma,a>m}^{\text{R}}~,
\end{align}
with $|n|+1$ boundary conditions, $I^\text{D3}_0=0$, $\sum_a t_a I_a^\text{S} = \nu_{-|n|} V$, and $I^\text{D1}_a=I^\text{D1}_b$ for $a,b\in[1,|n|]$. Similarly to the positive $n$ case, we rewrite these equations in a matrix form
\begin{align}
\begin{pmatrix}
    I^\text{D2}_\text{0}\\
    \vect{I}^\text{S}
    \end{pmatrix}
    =    W\hat{\cal P}_{m}W^{-1}
    \begin{pmatrix}
    I^\text{D3}_\text{0}\\
    \vect{I}^\text{D1}
    \end{pmatrix}.
\end{align}
Notice that, unlike the positive $n$ case, the projector $\hat{\cal P}_{m}$ acts on the $(I^\text{R}_\text{ch},\vect{I}^\text{R}_{\sigma})$ currents. We introduce $i_\text{D1}$ as the current per mode in the D1 lead so that the equilibration conditions at $x=x_R$ are resolved by setting $\vect{I}^\text{D1}=i_{D1} \vect{t}$. By multiplying the second set of equations for $\vect{I}^\text{S}$  by $\vect{t}^T$, we obtain
\begin{align}    
I^\text{D2}_\text{0}&=T_{00}I^\text{D3}_\text{0} + \vect{T}_{01}^T\vect{t} \;i_\text{D1}\\
I^\text{S}_\text{tot}&=\vect{t}^T\vect{T}_{10}I^\text{D3}_\text{0}+ \vect{t}^T\hat{T}_{11}\vect{t}\; i_\text{D1},
\end{align}
where the total current from the source is $I^\text{S}_\text{tot}=\vect{t}^T\vect{I}^S$. The product of matrices $ W\hat{\cal P}_{m}W^{-1}$ for negative $n$ is explicitly given by
\begin{align*}
 W\hat{\cal P}_{m}W^{-1}=&\begin{pmatrix}
    1-2m & \cdots & 1-2m & & -2m & \cdots & -2m \\
    2 & \cdots & 2 & & 2 & \cdots & 2 \\
    \vdots & & \vdots & & \vdots & & \vdots \\
    2 & \cdots & 2 & & 2 & \cdots & 2 \\
    0 & \cdots & 0 & & 1 & \cdots &0 \\
    \vdots & & \vdots & &  \vdots & \ddots &  \vdots \\
    0 & \cdots & 0 & &0 & \cdots & 1
\end{pmatrix}.\\
&\raisebox{1.2cm}{$\hspace{0.7cm}\underbrace{\hphantom{\hspace{2.3cm}}}_{\text{$m+1$ entries}}
\hspace{1.cm}
\underbrace{\hphantom{\hspace{1.9cm}}}_{\text{$|n|-m$ entries}}$}
\end{align*}
Assuming contacts S and D3 are at voltages $V_\text{S}$ and $V_\text{D3}$, respectively, we solve the equation for the outgoing currents to the drains D1 and D2, $I^\text{D1}_\text{tot}\equiv \vect{t}^T\vect{I}^\text{D1}=|n|i_\text{D1}$ and $I^\text{D2}_\text{0}$, and find
\begin{align}
    I^\text{D1}_\text{tot} &= \left(\frac{|n|}{2|n|-1} - \frac{ |n| m}{(2|n|-1)m+n}\right)V_\text{S} \nonumber\\&+ \;\frac{ |n| m}{(2|n|-1)m+|n|}V_\text{D3},\\
    -I^\text{D2}_\text{0} &= \frac{ |n| m}{(2|n|-1)m+n}V_\text{S} \nonumber \\&+\; \left(\frac{1}{2}-\frac{|n|m}{(2|n|-1)m+|n|}\right)V_\text{D3}.
\end{align}

Notice the close similarity to Eqs.~\eqref{eqn.sup.c1} and \eqref{eqn.sup.c2}, where the factor $\frac{m}{2m+1}$ is replaced by $\frac{nm}{(2n-1)m+n}$. Hence, for $V_\text{D3}=0$, we obtain the coherent conductance between the source S and the drain D1 in the $n<0$ case  
\begin{align}
    G_\ell=\frac{|n|}{2|n|-1} - \frac{ |n| m(\ell)}{(2|n|-1)m(\ell)+|n|},
\end{align}
which was used for Table~\ref{tab.condunctance} in the main text.

\subsection{Conductance of Anti-Read-Rezayi state}\label{app.anti_read}
The edge of the Read-Rezayi~\cite{Read_Beyond_1999} state at $\nu=\frac{2}{3}$ contains a chiral boson $\phi_1$ and neutral $\mathbb{Z}_3$ parafermion theory with central charge $c_{\mathbb{Z}_3}=\frac{4}{5}$.~\cite{Zamolodchikov_parafermion_1985}
The latter can be represented as an $SU(2)$ level-$3$ chiral Wess-Zumino-Witten model with gauged out $U(1)$ subgroup.~\cite{Karabali_WZW_1990,Witten_non_abelian_bosonization_1984,Bishara_PH_Read_Rezayi_2008} In accordance with Ref.~\onlinecite{Bishara_PH_Read_Rezayi_2008}, the particle-hole conjugate of the Read-Rezayi state contains the electron mode $\phi_2$ and the counter-propagating $\phi_1$ carrying $\frac{3}{5}$ charge conductance. The bosonic sector can be expressed in terms of the charge and the neutral modes governed by
\begin{align}\label{eq.app_LRR}
    {\cal L}_\rho &= \frac{1}{4\pi}\frac{5}{2}\partial_x\phi_\rho\left(i\partial_x\tau+ v_\rho\partial_x\right)\phi_\rho,\\
    {\cal L}_\sigma &= \frac{1}{4\pi}\frac{3}{2}\partial_x\phi_\sigma\left(i\partial_x\tau+ v_\sigma\partial_x\right)\phi_\sigma.
\end{align} 
The neutral mode $\phi_\sigma$, together with the chiral $\mathbb{Z}_3$ parafermion theory, can be combined to form a full $SU(2)_3$ Wess-Zumino-Witten model with central charge $c_{SU(2)_3} = \frac{9}{5}$.  

We assess the topological stability in the basis where the edge theory consists of $(\phi_1,\phi_2)=\frac{1}{2}(5\phi_\rho-3\phi_\sigma,3\phi_\rho-3\phi_\sigma)$ and $\mathbb{Z}_3$ parafermions. The $\mathbb{Z}_3$ theory has $c_{\mathbb{Z}_3}=\frac{4}{5}$ and cannot partially localize an Abelian boson mode, as the remaining theory would have $c=\frac{1}{5}$ and thus be non-unitary.~\cite{DiFrancesco_CFT_1997} Consequently, the $\mathbb{Z}_3$ sector does not participate in localization, and the analysis of the topological stability of bosons is carried out by means of Ref.~\onlinecite{Haldane_stability_1995}.

\subsubsection{The $\nu=1$ interface}
\begin{figure}[ht!]
    \centering
    \includegraphics[width=1\linewidth]{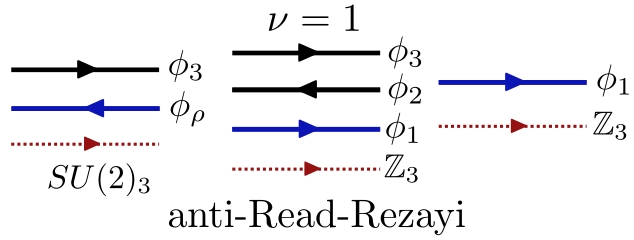}
    \caption{The edge structure of the interfaces between the $\nu=\frac{2}{5}$ anti-Read-Rezayi and $\nu=1$ integer quantum Hall states.}
    \label{fig.App_aRR_edge}
\end{figure}

We closely follow the analysis of Ref.~\onlinecite{Bishara_PH_Read_Rezayi_2008}. We add to the Lagrangian Eq.~\eqref{eq.app_LRR} an integer electron mode, $\phi_3$; see Fig.~\ref{fig.App_aRR_edge}. The counter-propagating electrons $\phi_2$ and $\phi_3$ localize each other if the constriction is longer than the localization length. After localization, the current $I_2$ is back-scattered into $I_3$, imposing 
\begin{align}\label{eq.app_loc}
    I^\text{S}_3=I^\text{D2}_2,\quad I^\text{D1}_3=I^\text{D3}_2.
\end{align}
Similarly to the negative $n<0$ case for paired states, equilibration in the leads implies that the modes $\phi_1$ and $\phi_2$ are in equilibrium at $D2$, which results in 
\begin{align}\label{eq.app_equil1}
    -\frac{5}{3}I^\text{D2}_{1}=I^\text{D2}_{2}.
\end{align}
In the coherent regime, there are no tunneling processes involving $\phi_1$ in the constriction, such that $I^\text{D2}_1=I^\text{D3}_1$. Including the boundary condition set by S and D3
\begin{align}\label{eq.app_equil2}
    I^\text{S}_3=V,\quad I^\text{D3}_1+I^\text{D3}_2=0,
\end{align}
we solve the equation to obtain $I^\text{D1}_3=\frac{3}{5}V$, which corresponds to fully equilibrated conductance $G=\frac{3}{5}$. This result is expected since, after the localization of two electrons, the edge in the constriction is chiral.

\subsubsection{The $\nu=2/3$ interface}
The $\nu=\frac{2}{3}$ interface differs from the previous case by a $\delta \nu=-\frac{1}{3}$ edge mode $\phi_4$. This extra mode does not allow any further localization, and the edge structure in the construction contains two counter-propagating modes, $\phi_1$ and $\phi_4$.

At the trivial fixed point, the modes $\phi_1$ and $\phi_4$ are continuous. Using the techniques of Appendix~\ref{app.Green}, we find that the most relevant tunneling operator is $\psi_1 e^{i\frac{5}{3}\phi_1+3i\phi_4}$, where parafermion field $\psi_1$ has scaling dimension $\frac{2}{3}$. We find that the scaling dimension of the tunneling operator $[\psi_1]+[e^{i\frac{5}{3}\phi_1}]+[e^{3i\phi_4}]=\frac{2}{3}+\frac{10}{3} + 3>1$, and the process is irrelevant.~\cite{Bishara_PH_Read_Rezayi_2008} Hence, the trivial fixed point is stable, and we proceed to the calculation of the coherent conductance.

The boundary condition set by the source is
\begin{align}
    I^\text{S}_{3}+I^\text{S}_4 = \frac{2}{3}V.
\end{align}
The equilibration of chemical potentials in the lead D1 reads
\begin{align}
    I^\text{D1}_{3}=-3I^\text{D1}_4.
\end{align}
For the same reason, the $I_4$ is also conserved in the construction $I^\text{D2}_4=I^\text{D3}_4$. The equations for currents $I_1$ and $I_2$ are the same as in the previous case. The solution yields the equal splitting of current between the source S and the drain D1, i.e., the conductance is $G=\frac{1}{3}$; see Table~\ref{tab.condunctance} in the main text. This result is particularly interesting as this value is inconsistent with fully equilibrated and ballistic regimes. Thus, a measurement of $G=\frac{1}{3}$ at these interfaces would imply $\mathbb{Z}_3$ parafermions topological order. 

\subsubsection{The $\nu=1/3$ interface}

The interface of anti-Read-Rezayi and the Laughlin state at $\nu=\frac{1}{3}$ is topologically stable since there is no tunneling operator between the two edges with zero charge and trivial self-statistics. The trivial fixed point is also stable in the renormalization group sense since the scaling dimension of $[\psi_1]+[e^{i\frac{5}{3}\phi_1}]+[e^{3i\phi_2}]=\frac{2}{3}+\frac{10}{3} + \frac{3}{2}>1 $. (Notice that, compared to the $\frac{2}{3}$ case, the $\phi_2$ mode of $\nu=\frac{1}{3}$ is decoupled and has a scaling dimension of $\frac{3}{2}$ instead of $3$.) Since there is no localization, the tunneling in the constriction is suppressed, and the coherent conductance assumes the ballistic value $G=\frac{1}{3}$.

\subsection{Conductance of Bonderson-Slingerland states}
The edge theory of the Bonderson-Slingerland state at $\nu=\frac{2}{5}$ is described by Eq.~\eqref{eq.LK} in the main text with
\begin{align}
    K_\text{BS}=\begin{pmatrix}
        2 & 3\\
        3 & 2\\
    \end{pmatrix},\quad \vect{t}=\begin{pmatrix}
        1\\
        1\\
    \end{pmatrix},
\end{align}
and ${\cal L}_{\ell=-3,1}$ in Eq.~\eqref{eq.Lgamma} for anti-Pfaffian and Moore-Read pairing channels.

\subsubsection{The $\nu=1$ interface}
This interface is described by $K=\text{diag}(1,-K_\text{BS})$, $t_a=1$, and ${\cal L}_{-\ell}$. An alternative basis $K'=W^TKW$ is obtained by a $W\in SL(3,\mathbb{Z})$ transformation with
\begin{align}
    K'=\begin{pmatrix}
        3 & 2 & 0\\
        2 & 3 & 0\\
        0 & 0 & -1
    \end{pmatrix},\quad
    W=\begin{pmatrix}
        1 & 1 & 1\\
        -1 & -1 & -2\\
        1 & 2 & 3
    \end{pmatrix},
\end{align}
and $\vect{t}'=W^T\vect{t}=(1,0,0)$. The mode $e^{i\phi'_3}$ represents a neutral upstream fermion. For the Bonderson-Slingerland state with anti-Pfaffian pairing, this fermion mode localizes with two $\gamma$ Majoranas. Consequently, the neutral current does not reach the left end of the junction, $I'_3(x_L)=0$, resulting in $G^3_\text{BS-aPf} = \frac{3}{5}$. For Moore-Read pairing, the edge is topologically stable since Majorana mode moves in the same direction as $\phi'_3$ mode, and the conductance is $G^3_\text{BS-MR}=1$. 

\subsubsection{The $\nu=2/3$ interface}
The $K$-matrix of this interface, $K=\text{diag}(-3,1,-K_\text{BS})$, can be transformed to $K'=\text{diag}(15,-1,-\sigma_x)$ with 
\begin{align}
    W=\begin{pmatrix}
        -1 & 0 & 0 & 1\\
        0 & 1 & 2 & 3\\
        3 & 1 & 1 & -1\\
        3 & -1 & -1 & -4
    \end{pmatrix}
\end{align}
and $\vect{t}'=(2,0,1,0)$. Similarly to the $\nu=1$ case, the neutral fermion $e^{i\phi'_2}$ localizes with $\gamma$ Majoranas depending on the pairing channel. In addition, there is a null-vector $\vect{m} = (3,1,1,-1)$ for any pairing channel. The current calculation yields $G^{8/3}_\text{BS-MR}=\frac{2}{3}$ and $G^{8/3}_\text{BS-aPf}=\frac{4}{15}$.

\subsubsection{The $\nu=1/3$ interface}
Finally, the interface $K=\text{diag}(3,-K_\text{BS})$ maps onto $K'=\text{diag}(-15,1,1)$ and $\vect{t}'=(1,0,0)$ with 
\begin{align}
    W=\begin{pmatrix}
        1 & 1 & 1 \\
        3 & 3 & 2 \\
        3 & 2 & 3 
    \end{pmatrix}.
\end{align}
For Moore-Read pairing, the Majorana mode localizes with one Majorana of $e^{i\phi'_{2}}$ or $e^{i\phi'_{3}}$ and imposes $I'_{2}(x_R)=0$ or $I'_{2}(x_R)=0$, respectively. The resulting coherent conductance is $G^{7/3}_\text{BS-MR} = \frac{1}{21}$. Meanwhile, the interface is topologically stable for anti-Pfaffian pairing and exhibits $G^{7/3}_\text{BS-aPf} = \frac{1}{3}$.

\bibliography{ref}

%merlin.mbs apsrev4-1.bst 2010-07-25 4.21a (PWD, AO, DPC) hacked
%Control: key (0)
%Control: author (0) dotless jnrlst
%Control: editor formatted (1) identically to author
%Control: production of article title (0) allowed
%Control: page (1) range
%Control: year (0) verbatim
%Control: production of eprint (0) enabled
\begin{thebibliography}{69}%
\makeatletter
\providecommand \@ifxundefined [1]{%
 \@ifx{#1\undefined}
}%
\providecommand \@ifnum [1]{%
 \ifnum #1\expandafter \@firstoftwo
 \else \expandafter \@secondoftwo
 \fi
}%
\providecommand \@ifx [1]{%
 \ifx #1\expandafter \@firstoftwo
 \else \expandafter \@secondoftwo
 \fi
}%
\providecommand \natexlab [1]{#1}%
\providecommand \enquote  [1]{``#1''}%
\providecommand \bibnamefont  [1]{#1}%
\providecommand \bibfnamefont [1]{#1}%
\providecommand \citenamefont [1]{#1}%
\providecommand \href@noop [0]{\@secondoftwo}%
\providecommand \href [0]{\begingroup \@sanitize@url \@href}%
\providecommand \@href[1]{\@@startlink{#1}\@@href}%
\providecommand \@@href[1]{\endgroup#1\@@endlink}%
\providecommand \@sanitize@url [0]{\catcode `\\12\catcode `\$12\catcode `\&12\catcode `\#12\catcode `\^12\catcode `\_12\catcode `\%12\relax}%
\providecommand \@@startlink[1]{}%
\providecommand \@@endlink[0]{}%
\providecommand \url  [0]{\begingroup\@sanitize@url \@url }%
\providecommand \@url [1]{\endgroup\@href {#1}{\urlprefix }}%
\providecommand \urlprefix  [0]{URL }%
\providecommand \Eprint [0]{\href }%
\providecommand \doibase [0]{http://dx.doi.org/}%
\providecommand \selectlanguage [0]{\@gobble}%
\providecommand \bibinfo  [0]{\@secondoftwo}%
\providecommand \bibfield  [0]{\@secondoftwo}%
\providecommand \translation [1]{[#1]}%
\providecommand \BibitemOpen [0]{}%
\providecommand \bibitemStop [0]{}%
\providecommand \bibitemNoStop [0]{.\EOS\space}%
\providecommand \EOS [0]{\spacefactor3000\relax}%
\providecommand \BibitemShut  [1]{\csname bibitem#1\endcsname}%
\let\auto@bib@innerbib\@empty
%</preamble>
\bibitem [{\citenamefont {Tsui}\ \emph {et~al.}(1982)\citenamefont {Tsui}, \citenamefont {Stormer},\ and\ \citenamefont {Gossard}}]{Tsui_fqh_1982}%
  \BibitemOpen
  \bibfield  {author} {\bibinfo {author} {\bibfnamefont {D.~C.}\ \bibnamefont {Tsui}}, \bibinfo {author} {\bibfnamefont {H.~L.}\ \bibnamefont {Stormer}}, \ and\ \bibinfo {author} {\bibfnamefont {A.~C.}\ \bibnamefont {Gossard}},\ }\bibfield  {title} {\enquote {\bibinfo {title} {Two-dimensional magnetotransport in the extreme quantum limit},}\ }\href {\doibase 10.1103/PhysRevLett.48.1559} {\bibfield  {journal} {\bibinfo  {journal} {Physical Review Letters}\ }\textbf {\bibinfo {volume} {48}},\ \bibinfo {pages} {1559} (\bibinfo {year} {1982})}\BibitemShut {NoStop}%
\bibitem [{\citenamefont {Laughlin}(1983)}]{Laughlin_fqh_1983}%
  \BibitemOpen
  \bibfield  {author} {\bibinfo {author} {\bibfnamefont {R.~B.}\ \bibnamefont {Laughlin}},\ }\bibfield  {title} {\enquote {\bibinfo {title} {Anomalous quantum {Hall} effect: An incompressible quantum fluid with fractionally charged excitations},}\ }\href {\doibase 10.1103/PhysRevLett.50.1395} {\bibfield  {journal} {\bibinfo  {journal} {Physical Review Letters}\ }\textbf {\bibinfo {volume} {50}},\ \bibinfo {pages} {1395} (\bibinfo {year} {1983})}\BibitemShut {NoStop}%
\bibitem [{\citenamefont {Haldane}(1983)}]{Haldane_fqh_1983}%
  \BibitemOpen
  \bibfield  {author} {\bibinfo {author} {\bibfnamefont {F.~D.~M.}\ \bibnamefont {Haldane}},\ }\bibfield  {title} {\enquote {\bibinfo {title} {Fractional quantization of the {Hall} effect: {A} hierarchy of incompressible quantum fluid states},}\ }\href {https://journals.aps.org/prl/abstract/10.1103/PhysRevLett.51.605} {\bibfield  {journal} {\bibinfo  {journal} {Physical Review Letters}\ }\textbf {\bibinfo {volume} {51}},\ \bibinfo {pages} {605} (\bibinfo {year} {1983})}\BibitemShut {NoStop}%
\bibitem [{\citenamefont {Rammal}\ \emph {et~al.}(1983)\citenamefont {Rammal}, \citenamefont {Toulouse}, \citenamefont {Jaekel},\ and\ \citenamefont {Halperin}}]{Halperin_fqh_1983}%
  \BibitemOpen
  \bibfield  {author} {\bibinfo {author} {\bibfnamefont {R.}~\bibnamefont {Rammal}}, \bibinfo {author} {\bibfnamefont {G.}~\bibnamefont {Toulouse}}, \bibinfo {author} {\bibfnamefont {M.~T.}\ \bibnamefont {Jaekel}}, \ and\ \bibinfo {author} {\bibfnamefont {B.~I.}\ \bibnamefont {Halperin}},\ }\bibfield  {title} {\enquote {\bibinfo {title} {Quantized {Hall} conductance and edge states: Two-dimensional strips with a periodic potential},}\ }\href {\doibase 10.1103/PhysRevB.27.5142} {\bibfield  {journal} {\bibinfo  {journal} {Physical Review B}\ }\textbf {\bibinfo {volume} {27}},\ \bibinfo {pages} {5142} (\bibinfo {year} {1983})}\BibitemShut {NoStop}%
\bibitem [{\citenamefont {Wen}(1991{\natexlab{a}})}]{Wen_Non-Abelian_1991}%
  \BibitemOpen
  \bibfield  {author} {\bibinfo {author} {\bibfnamefont {X.~G.}\ \bibnamefont {Wen}},\ }\bibfield  {title} {\enquote {\bibinfo {title} {Non-{Abelian} statistics in the fractional quantum {Hall} states},}\ }\href {\doibase 10.1103/PhysRevLett.66.802} {\bibfield  {journal} {\bibinfo  {journal} {Physical Review Letters}\ }\textbf {\bibinfo {volume} {66}},\ \bibinfo {pages} {802--805} (\bibinfo {year} {1991}{\natexlab{a}})}\BibitemShut {NoStop}%
\bibitem [{\citenamefont {Moore}\ and\ \citenamefont {Read}(1991)}]{Moore_nonabelions_1991}%
  \BibitemOpen
  \bibfield  {author} {\bibinfo {author} {\bibfnamefont {G.}~\bibnamefont {Moore}}\ and\ \bibinfo {author} {\bibfnamefont {N.}~\bibnamefont {Read}},\ }\bibfield  {title} {\enquote {\bibinfo {title} {Nonabelions in the fractional quantum {Hall} effect},}\ }\href {http://www.sciencedirect.com/science/article/pii/055032139190407O} {\bibfield  {journal} {\bibinfo  {journal} {Nucl. Phys. B}\ }\textbf {\bibinfo {volume} {360}},\ \bibinfo {pages} {362} (\bibinfo {year} {1991})}\BibitemShut {NoStop}%
\bibitem [{\citenamefont {Read}\ and\ \citenamefont {Green}(2000)}]{Read_paired_2000}%
  \BibitemOpen
  \bibfield  {author} {\bibinfo {author} {\bibfnamefont {N.}~\bibnamefont {Read}}\ and\ \bibinfo {author} {\bibfnamefont {D.}~\bibnamefont {Green}},\ }\bibfield  {title} {\enquote {\bibinfo {title} {Paired states of fermions in two dimensions with breaking of parity and time-reversal symmetries and the fractional quantum {Hall} effect},}\ }\href {https://journals.aps.org/prb/abstract/10.1103/PhysRevB.61.10267} {\bibfield  {journal} {\bibinfo  {journal} {Physical Review B}\ }\textbf {\bibinfo {volume} {61}},\ \bibinfo {pages} {10267} (\bibinfo {year} {2000})}\BibitemShut {NoStop}%
\bibitem [{\citenamefont {Stern}\ and\ \citenamefont {Halperin}(2006)}]{Stern_Probe_Non_Abelian_2006}%
  \BibitemOpen
  \bibfield  {author} {\bibinfo {author} {\bibfnamefont {A.}~\bibnamefont {Stern}}\ and\ \bibinfo {author} {\bibfnamefont {B.~I.}\ \bibnamefont {Halperin}},\ }\bibfield  {title} {\enquote {\bibinfo {title} {Proposed experiments to probe the non-{Abelian} $\ensuremath{\nu}=5/2$ quantum {Hall} state},}\ }\href {\doibase 10.1103/PhysRevLett.96.016802} {\bibfield  {journal} {\bibinfo  {journal} {Physical Review Letters}\ }\textbf {\bibinfo {volume} {96}},\ \bibinfo {pages} {016802} (\bibinfo {year} {2006})}\BibitemShut {NoStop}%
\bibitem [{\citenamefont {Nayak}\ \emph {et~al.}(2008)\citenamefont {Nayak}, \citenamefont {Simon}, \citenamefont {Stern}, \citenamefont {Freedman},\ and\ \citenamefont {Das~Sarma}}]{nayak_non-abelian_2008}%
  \BibitemOpen
  \bibfield  {author} {\bibinfo {author} {\bibfnamefont {C.}~\bibnamefont {Nayak}}, \bibinfo {author} {\bibfnamefont {S.~H.}\ \bibnamefont {Simon}}, \bibinfo {author} {\bibfnamefont {A.}~\bibnamefont {Stern}}, \bibinfo {author} {\bibfnamefont {M.}~\bibnamefont {Freedman}}, \ and\ \bibinfo {author} {\bibfnamefont {S.}~\bibnamefont {Das~Sarma}},\ }\bibfield  {title} {\enquote {\bibinfo {title} {Non-{Abelian} anyons and topological quantum computation},}\ }\href {\doibase 10.1103/RevModPhys.80.1083} {\bibfield  {journal} {\bibinfo  {journal} {Rev. Mod. Phys.}\ }\textbf {\bibinfo {volume} {80}},\ \bibinfo {pages} {1083} (\bibinfo {year} {2008})}\BibitemShut {NoStop}%
\bibitem [{\citenamefont {{Stern}}(2010)}]{Stern_non_Abelian_2010}%
  \BibitemOpen
  \bibfield  {author} {\bibinfo {author} {\bibfnamefont {A.}~\bibnamefont {{Stern}}},\ }\bibfield  {title} {\enquote {\bibinfo {title} {{Non-Abelian states of matter}},}\ }\href {\doibase 10.1038/nature08915} {\bibfield  {journal} {\bibinfo  {journal} {\nat}\ }\textbf {\bibinfo {volume} {464}},\ \bibinfo {pages} {187--193} (\bibinfo {year} {2010})}\BibitemShut {NoStop}%
\bibitem [{\citenamefont {Freedman}\ \emph {et~al.}(2000)\citenamefont {Freedman}, \citenamefont {Larsen},\ and\ \citenamefont {Wang}}]{Freedman_modular_functor_2000}%
  \BibitemOpen
  \bibfield  {author} {\bibinfo {author} {\bibfnamefont {M.}~\bibnamefont {Freedman}}, \bibinfo {author} {\bibfnamefont {M.}~\bibnamefont {Larsen}}, \ and\ \bibinfo {author} {\bibfnamefont {Z.}~\bibnamefont {Wang}},\ }\href@noop {} {\enquote {\bibinfo {title} {A modular functor which is universal for quantum computation},}\ } (\bibinfo {year} {2000}),\ \Eprint {http://arxiv.org/abs/quant-ph/0001108} {arXiv:quant-ph/0001108 [quant-ph]} \BibitemShut {NoStop}%
\bibitem [{\citenamefont {Bonesteel}\ \emph {et~al.}(2005)\citenamefont {Bonesteel}, \citenamefont {Hormozi}, \citenamefont {Zikos},\ and\ \citenamefont {Simon}}]{Bonesteel_Braid_Topologies_2005}%
  \BibitemOpen
  \bibfield  {author} {\bibinfo {author} {\bibfnamefont {N.~E.}\ \bibnamefont {Bonesteel}}, \bibinfo {author} {\bibfnamefont {L.}~\bibnamefont {Hormozi}}, \bibinfo {author} {\bibfnamefont {G.}~\bibnamefont {Zikos}}, \ and\ \bibinfo {author} {\bibfnamefont {S.~H.}\ \bibnamefont {Simon}},\ }\bibfield  {title} {\enquote {\bibinfo {title} {Braid topologies for quantum computation},}\ }\href {\doibase 10.1103/PhysRevLett.95.140503} {\bibfield  {journal} {\bibinfo  {journal} {Physical Review Letters}\ }\textbf {\bibinfo {volume} {95}},\ \bibinfo {pages} {140503} (\bibinfo {year} {2005})}\BibitemShut {NoStop}%
\bibitem [{\citenamefont {Hormozi}\ \emph {et~al.}(2007)\citenamefont {Hormozi}, \citenamefont {Zikos}, \citenamefont {Bonesteel},\ and\ \citenamefont {Simon}}]{Hormozi_Topological_quantum_2007}%
  \BibitemOpen
  \bibfield  {author} {\bibinfo {author} {\bibfnamefont {L.}~\bibnamefont {Hormozi}}, \bibinfo {author} {\bibfnamefont {G.}~\bibnamefont {Zikos}}, \bibinfo {author} {\bibfnamefont {N.~E.}\ \bibnamefont {Bonesteel}}, \ and\ \bibinfo {author} {\bibfnamefont {S.~H.}\ \bibnamefont {Simon}},\ }\bibfield  {title} {\enquote {\bibinfo {title} {Topological quantum compiling},}\ }\href {\doibase 10.1103/PhysRevB.75.165310} {\bibfield  {journal} {\bibinfo  {journal} {Physical Review B}\ }\textbf {\bibinfo {volume} {75}},\ \bibinfo {pages} {165310} (\bibinfo {year} {2007})}\BibitemShut {NoStop}%
\bibitem [{\citenamefont {Willett}\ \emph {et~al.}(1987)\citenamefont {Willett}, \citenamefont {Eisenstein}, \citenamefont {St\"ormer}, \citenamefont {Tsui}, \citenamefont {Gossard},\ and\ \citenamefont {English}}]{Willett_observation_1987}%
  \BibitemOpen
  \bibfield  {author} {\bibinfo {author} {\bibfnamefont {R.}~\bibnamefont {Willett}}, \bibinfo {author} {\bibfnamefont {J.~P.}\ \bibnamefont {Eisenstein}}, \bibinfo {author} {\bibfnamefont {H.~L.}\ \bibnamefont {St\"ormer}}, \bibinfo {author} {\bibfnamefont {D.~C.}\ \bibnamefont {Tsui}}, \bibinfo {author} {\bibfnamefont {A.~C.}\ \bibnamefont {Gossard}}, \ and\ \bibinfo {author} {\bibfnamefont {J.~H.}\ \bibnamefont {English}},\ }\bibfield  {title} {\enquote {\bibinfo {title} {Observation of an even-denominator quantum number in the fractional quantum {Hall} effect},}\ }\href {\doibase 10.1103/PhysRevLett.59.1776} {\bibfield  {journal} {\bibinfo  {journal} {Physical Review Letters}\ }\textbf {\bibinfo {volume} {59}},\ \bibinfo {pages} {1776} (\bibinfo {year} {1987})}\BibitemShut {NoStop}%
\bibitem [{\citenamefont {Kim}\ \emph {et~al.}(2019)\citenamefont {Kim}, \citenamefont {Balram}, \citenamefont {Taniguchi}, \citenamefont {Watanabe}, \citenamefont {Jain},\ and\ \citenamefont {Smet}}]{Kim_Even_Denominator_f_wave_2019}%
  \BibitemOpen
  \bibfield  {author} {\bibinfo {author} {\bibfnamefont {Y.}~\bibnamefont {Kim}}, \bibinfo {author} {\bibfnamefont {A.~C.}\ \bibnamefont {Balram}}, \bibinfo {author} {\bibfnamefont {T.}~\bibnamefont {Taniguchi}}, \bibinfo {author} {\bibfnamefont {K.}~\bibnamefont {Watanabe}}, \bibinfo {author} {\bibfnamefont {J.~K.}\ \bibnamefont {Jain}}, \ and\ \bibinfo {author} {\bibfnamefont {J.~H.}\ \bibnamefont {Smet}},\ }\bibfield  {title} {\enquote {\bibinfo {title} {Even-denominator fractional quantum {Hall} states in higher {Landau} levels of graphene},}\ }\href {\doibase 10.1038/s41567-018-0355-x} {\bibfield  {journal} {\bibinfo  {journal} {Nature Physics}\ }\textbf {\bibinfo {volume} {15}},\ \bibinfo {pages} {154--158} (\bibinfo {year} {2019})}\BibitemShut {NoStop}%
\bibitem [{\citenamefont {Ki}\ \emph {et~al.}(2014)\citenamefont {Ki}, \citenamefont {Fal’ko}, \citenamefont {Abanin},\ and\ \citenamefont {Morpurgo}}]{Ki_bilyaer_graphene_2014}%
  \BibitemOpen
  \bibfield  {author} {\bibinfo {author} {\bibfnamefont {D.~K.}\ \bibnamefont {Ki}}, \bibinfo {author} {\bibfnamefont {V.~I.}\ \bibnamefont {Fal’ko}}, \bibinfo {author} {\bibfnamefont {D.~A.}\ \bibnamefont {Abanin}}, \ and\ \bibinfo {author} {\bibfnamefont {A.~F.}\ \bibnamefont {Morpurgo}},\ }\bibfield  {title} {\enquote {\bibinfo {title} {Observation of even denominator fractional quantum {Hall} effect in suspended bilayer graphene},}\ }\href {\doibase 10.1021/nl5003922} {\bibfield  {journal} {\bibinfo  {journal} {Nano Letters}\ }\textbf {\bibinfo {volume} {14}},\ \bibinfo {pages} {2135--2139} (\bibinfo {year} {2014})}\BibitemShut {NoStop}%
\bibitem [{\citenamefont {Kim}\ \emph {et~al.}(2015)\citenamefont {Kim}, \citenamefont {Lee}, \citenamefont {Jung}, \citenamefont {Skákalová}, \citenamefont {Taniguchi}, \citenamefont {Watanabe}, \citenamefont {Kim},\ and\ \citenamefont {Smet}}]{Kim_bilayer_graphene_2015}%
  \BibitemOpen
  \bibfield  {author} {\bibinfo {author} {\bibfnamefont {Y.}~\bibnamefont {Kim}}, \bibinfo {author} {\bibfnamefont {D.~S.}\ \bibnamefont {Lee}}, \bibinfo {author} {\bibfnamefont {S.}~\bibnamefont {Jung}}, \bibinfo {author} {\bibfnamefont {V.}~\bibnamefont {Skákalová}}, \bibinfo {author} {\bibfnamefont {T.}~\bibnamefont {Taniguchi}}, \bibinfo {author} {\bibfnamefont {K.}~\bibnamefont {Watanabe}}, \bibinfo {author} {\bibfnamefont {J.~S.}\ \bibnamefont {Kim}}, \ and\ \bibinfo {author} {\bibfnamefont {J.~H.}\ \bibnamefont {Smet}},\ }\bibfield  {title} {\enquote {\bibinfo {title} {Fractional quantum {Hall} states in bilayer graphene probed by transconductance fluctuations},}\ }\href {\doibase 10.1021/acs.nanolett.5b02876} {\bibfield  {journal} {\bibinfo  {journal} {Nano Letters}\ }\textbf {\bibinfo {volume} {15}},\ \bibinfo {pages} {7445--7451} (\bibinfo {year} {2015})}\BibitemShut {NoStop}%
\bibitem [{\citenamefont {Li}\ \emph {et~al.}(2017)\citenamefont {Li}, \citenamefont {Tan}, \citenamefont {Chen}, \citenamefont {Zeng}, \citenamefont {Taniguchi}, \citenamefont {Watanabe}, \citenamefont {Hone},\ and\ \citenamefont {Dean}}]{Li_bilayer_graphene_2017}%
  \BibitemOpen
  \bibfield  {author} {\bibinfo {author} {\bibfnamefont {J.~I.~A.}\ \bibnamefont {Li}}, \bibinfo {author} {\bibfnamefont {C.}~\bibnamefont {Tan}}, \bibinfo {author} {\bibfnamefont {S.}~\bibnamefont {Chen}}, \bibinfo {author} {\bibfnamefont {Y.}~\bibnamefont {Zeng}}, \bibinfo {author} {\bibfnamefont {T.}~\bibnamefont {Taniguchi}}, \bibinfo {author} {\bibfnamefont {K.}~\bibnamefont {Watanabe}}, \bibinfo {author} {\bibfnamefont {J.}~\bibnamefont {Hone}}, \ and\ \bibinfo {author} {\bibfnamefont {C.~R.}\ \bibnamefont {Dean}},\ }\bibfield  {title} {\enquote {\bibinfo {title} {Even-denominator fractional quantum {Hall} states in bilayer graphene},}\ }\href {\doibase 10.1126/science.aao2521} {\bibfield  {journal} {\bibinfo  {journal} {Science}\ }\textbf {\bibinfo {volume} {358}},\ \bibinfo {pages} {648--652} (\bibinfo {year} {2017})}\BibitemShut {NoStop}%
\bibitem [{\citenamefont {Zibrov}\ \emph {et~al.}(2017)\citenamefont {Zibrov}, \citenamefont {Kometter}, \citenamefont {Zhou}, \citenamefont {Spanton}, \citenamefont {Taniguchi}, \citenamefont {Watanabe}, \citenamefont {Zaletel},\ and\ \citenamefont {Young}}]{Zibrov_Tunable_bilayer_graphene_2017}%
  \BibitemOpen
  \bibfield  {author} {\bibinfo {author} {\bibfnamefont {A.~A.}\ \bibnamefont {Zibrov}}, \bibinfo {author} {\bibfnamefont {C.}~\bibnamefont {Kometter}}, \bibinfo {author} {\bibfnamefont {H.}~\bibnamefont {Zhou}}, \bibinfo {author} {\bibfnamefont {E.~M.}\ \bibnamefont {Spanton}}, \bibinfo {author} {\bibfnamefont {T.}~\bibnamefont {Taniguchi}}, \bibinfo {author} {\bibfnamefont {K.}~\bibnamefont {Watanabe}}, \bibinfo {author} {\bibfnamefont {M.~P.}\ \bibnamefont {Zaletel}}, \ and\ \bibinfo {author} {\bibfnamefont {A.~F.}\ \bibnamefont {Young}},\ }\bibfield  {title} {\enquote {\bibinfo {title} {Tunable interacting composite fermion phases in a half-filled bilayer-graphene {Landau} level},}\ }\href {\doibase 10.1038/nature23893} {\bibfield  {journal} {\bibinfo  {journal} {Nature}\ }\textbf {\bibinfo {volume} {549}},\ \bibinfo {pages} {360--364} (\bibinfo {year} {2017})}\BibitemShut {NoStop}%
\bibitem [{\citenamefont {Assouline}\ \emph {et~al.}(2024)\citenamefont {Assouline}, \citenamefont {Wang}, \citenamefont {Zhou}, \citenamefont {Cohen}, \citenamefont {Yang}, \citenamefont {Zhang}, \citenamefont {Taniguchi}, \citenamefont {Watanabe}, \citenamefont {Mong}, \citenamefont {Zaletel},\ and\ \citenamefont {Young}}]{Assouline_Energy_Gap_bilayer_graphene_2024}%
  \BibitemOpen
  \bibfield  {author} {\bibinfo {author} {\bibfnamefont {A.}~\bibnamefont {Assouline}}, \bibinfo {author} {\bibfnamefont {T.}~\bibnamefont {Wang}}, \bibinfo {author} {\bibfnamefont {H.}~\bibnamefont {Zhou}}, \bibinfo {author} {\bibfnamefont {L.~A.}\ \bibnamefont {Cohen}}, \bibinfo {author} {\bibfnamefont {F.}~\bibnamefont {Yang}}, \bibinfo {author} {\bibfnamefont {R.}~\bibnamefont {Zhang}}, \bibinfo {author} {\bibfnamefont {T.}~\bibnamefont {Taniguchi}}, \bibinfo {author} {\bibfnamefont {K.}~\bibnamefont {Watanabe}}, \bibinfo {author} {\bibfnamefont {R.~S.~K.}\ \bibnamefont {Mong}}, \bibinfo {author} {\bibfnamefont {M.~P.}\ \bibnamefont {Zaletel}}, \ and\ \bibinfo {author} {\bibfnamefont {A.~F.}\ \bibnamefont {Young}},\ }\bibfield  {title} {\enquote {\bibinfo {title} {Energy gap of the even-denominator fractional quantum hall state in bilayer graphene},}\ }\href {\doibase 10.1103/PhysRevLett.132.046603} {\bibfield  {journal} {\bibinfo  {journal} {Physical Review Letters}\ }\textbf {\bibinfo {volume} {132}},\
  \bibinfo {pages} {046603} (\bibinfo {year} {2024})}\BibitemShut {NoStop}%
\bibitem [{\citenamefont {{Kumar}}\ \emph {et~al.}(2024)\citenamefont {{Kumar}}, \citenamefont {{Haug}}, \citenamefont {{Kim}}, \citenamefont {{Yutushui}}, \citenamefont {{Khudiakov}}, \citenamefont {{Bhardwaj}}, \citenamefont {{Ilin}}, \citenamefont {{Watanabe}}, \citenamefont {{Taniguchi}}, \citenamefont {{Mross}},\ and\ \citenamefont {{Ronen}}}]{Kumar_Quarter_2024}%
  \BibitemOpen
  \bibfield  {author} {\bibinfo {author} {\bibfnamefont {R.}~\bibnamefont {{Kumar}}}, \bibinfo {author} {\bibfnamefont {A.}~\bibnamefont {{Haug}}}, \bibinfo {author} {\bibfnamefont {J.}~\bibnamefont {{Kim}}}, \bibinfo {author} {\bibfnamefont {M.}~\bibnamefont {{Yutushui}}}, \bibinfo {author} {\bibfnamefont {K.}~\bibnamefont {{Khudiakov}}}, \bibinfo {author} {\bibfnamefont {V.}~\bibnamefont {{Bhardwaj}}}, \bibinfo {author} {\bibfnamefont {A.}~\bibnamefont {{Ilin}}}, \bibinfo {author} {\bibfnamefont {K.}~\bibnamefont {{Watanabe}}}, \bibinfo {author} {\bibfnamefont {T.}~\bibnamefont {{Taniguchi}}}, \bibinfo {author} {\bibfnamefont {D.~F.}\ \bibnamefont {{Mross}}}, \ and\ \bibinfo {author} {\bibfnamefont {Y.}~\bibnamefont {{Ronen}}},\ }\bibfield  {title} {\enquote {\bibinfo {title} {{Quarter- and half-filled quantum Hall states and their competing interactions in bilayer graphene}},}\ }\href@noop {} {\bibfield  {journal} {\bibinfo  {journal} {arXiv e-prints}\ } (\bibinfo {year} {2024})},\ \Eprint
  {http://arxiv.org/abs/2405.19405} {arXiv:2405.19405 [cond-mat.mes-hall]} \BibitemShut {NoStop}%
\bibitem [{\citenamefont {Chen}\ \emph {et~al.}(2023)\citenamefont {Chen}, \citenamefont {Huang}, \citenamefont {Li}, \citenamefont {Tong}, \citenamefont {Kuang}, \citenamefont {Xi}, \citenamefont {Watanabe}, \citenamefont {Taniguchi}, \citenamefont {Liu}, \citenamefont {Zhu}, \citenamefont {Lu}, \citenamefont {Zhang}, \citenamefont {Wu},\ and\ \citenamefont {Wang}}]{chen_tunable_2023}%
  \BibitemOpen
  \bibfield  {author} {\bibinfo {author} {\bibfnamefont {Y.}~\bibnamefont {Chen}}, \bibinfo {author} {\bibfnamefont {Y.}~\bibnamefont {Huang}}, \bibinfo {author} {\bibfnamefont {Q.}~\bibnamefont {Li}}, \bibinfo {author} {\bibfnamefont {B.}~\bibnamefont {Tong}}, \bibinfo {author} {\bibfnamefont {G.}~\bibnamefont {Kuang}}, \bibinfo {author} {\bibfnamefont {C.}~\bibnamefont {Xi}}, \bibinfo {author} {\bibfnamefont {K.}~\bibnamefont {Watanabe}}, \bibinfo {author} {\bibfnamefont {T.}~\bibnamefont {Taniguchi}}, \bibinfo {author} {\bibfnamefont {G.}~\bibnamefont {Liu}}, \bibinfo {author} {\bibfnamefont {Z.}~\bibnamefont {Zhu}}, \bibinfo {author} {\bibfnamefont {L.}~\bibnamefont {Lu}}, \bibinfo {author} {\bibfnamefont {F.-C.}\ \bibnamefont {Zhang}}, \bibinfo {author} {\bibfnamefont {Y.-H.}\ \bibnamefont {Wu}}, \ and\ \bibinfo {author} {\bibfnamefont {L.}~\bibnamefont {Wang}},\ }\bibfield  {title} {\enquote {\bibinfo {title} {Tunable even-and odd-denominator fractional quantum hall states in trilayer graphene},}\
  }\href@noop {} {\  (\bibinfo {year} {2023})},\ \Eprint {http://arxiv.org/abs/2312.17204} {arXiv:2312.17204 [cond-mat.mes-hall]} \BibitemShut {NoStop}%
\bibitem [{\citenamefont {Falson}\ \emph {et~al.}(2015)\citenamefont {Falson}, \citenamefont {Maryenko}, \citenamefont {Friess}, \citenamefont {Zhang}, \citenamefont {Kozuka}, \citenamefont {Tsukazaki}, \citenamefont {Smet},\ and\ \citenamefont {Kawasaki}}]{Falson_Zno_2015}%
  \BibitemOpen
  \bibfield  {author} {\bibinfo {author} {\bibfnamefont {J.}~\bibnamefont {Falson}}, \bibinfo {author} {\bibfnamefont {D.}~\bibnamefont {Maryenko}}, \bibinfo {author} {\bibfnamefont {B.}~\bibnamefont {Friess}}, \bibinfo {author} {\bibfnamefont {D.}~\bibnamefont {Zhang}}, \bibinfo {author} {\bibfnamefont {Y.}~\bibnamefont {Kozuka}}, \bibinfo {author} {\bibfnamefont {A.}~\bibnamefont {Tsukazaki}}, \bibinfo {author} {\bibfnamefont {J.~H.}\ \bibnamefont {Smet}}, \ and\ \bibinfo {author} {\bibfnamefont {M.}~\bibnamefont {Kawasaki}},\ }\bibfield  {title} {\enquote {\bibinfo {title} {Even-denominator fractional quantum {Hall} physics in {ZnO}},}\ }\href {\doibase 10.1038/nphys3259} {\bibfield  {journal} {\bibinfo  {journal} {Nature Physics}\ }\textbf {\bibinfo {volume} {11}},\ \bibinfo {pages} {347--351} (\bibinfo {year} {2015})}\BibitemShut {NoStop}%
\bibitem [{\citenamefont {Falson}\ \emph {et~al.}(2018)\citenamefont {Falson}, \citenamefont {Tabrea}, \citenamefont {Zhang}, \citenamefont {Sodemann}, \citenamefont {Kozuka}, \citenamefont {Tsukazaki}, \citenamefont {Kawasaki}, \citenamefont {v.~Klitzing},\ and\ \citenamefont {Smet}}]{Falson_Zno_2018}%
  \BibitemOpen
  \bibfield  {author} {\bibinfo {author} {\bibfnamefont {J.}~\bibnamefont {Falson}}, \bibinfo {author} {\bibfnamefont {D.}~\bibnamefont {Tabrea}}, \bibinfo {author} {\bibfnamefont {D.}~\bibnamefont {Zhang}}, \bibinfo {author} {\bibfnamefont {I.}~\bibnamefont {Sodemann}}, \bibinfo {author} {\bibfnamefont {Y.}~\bibnamefont {Kozuka}}, \bibinfo {author} {\bibfnamefont {A.}~\bibnamefont {Tsukazaki}}, \bibinfo {author} {\bibfnamefont {M.}~\bibnamefont {Kawasaki}}, \bibinfo {author} {\bibfnamefont {K.}~\bibnamefont {v.~Klitzing}}, \ and\ \bibinfo {author} {\bibfnamefont {J.~H.}\ \bibnamefont {Smet}},\ }\bibfield  {title} {\enquote {\bibinfo {title} {A cascade of phase transitions in an orbitally mixed half-filled {Landau} level},}\ }\href {\doibase 10.1126/sciadv.aat8742} {\bibfield  {journal} {\bibinfo  {journal} {Science Advances}\ }\textbf {\bibinfo {volume} {4}},\ \bibinfo {pages} {eaat8742} (\bibinfo {year} {2018})}\BibitemShut {NoStop}%
\bibitem [{\citenamefont {Morf}(1998)}]{Morf_transition_1998}%
  \BibitemOpen
  \bibfield  {author} {\bibinfo {author} {\bibfnamefont {R.~H.}\ \bibnamefont {Morf}},\ }\bibfield  {title} {\enquote {\bibinfo {title} {Transition from quantum {Hall} to compressible states in the second {Landau} level: new light on the $\nu=$~5/2 enigma},}\ }\href {https://journals.aps.org/prl/abstract/10.1103/PhysRevLett.80.1505} {\bibfield  {journal} {\bibinfo  {journal} {Physical Review Letters}\ }\textbf {\bibinfo {volume} {80}},\ \bibinfo {pages} {1505} (\bibinfo {year} {1998})}\BibitemShut {NoStop}%
\bibitem [{\citenamefont {Rezayi}\ and\ \citenamefont {Haldane}(2000)}]{Rezayi_incompressible_2000}%
  \BibitemOpen
  \bibfield  {author} {\bibinfo {author} {\bibfnamefont {E.~H.}\ \bibnamefont {Rezayi}}\ and\ \bibinfo {author} {\bibfnamefont {F.~D.~M.}\ \bibnamefont {Haldane}},\ }\bibfield  {title} {\enquote {\bibinfo {title} {Incompressible paired {Hall} state, stripe order, and the composite fermion liquid phase in half-filled {Landau} levels},}\ }\href {https://journals.aps.org/prl/abstract/10.1103/PhysRevLett.84.4685} {\bibfield  {journal} {\bibinfo  {journal} {Physical Review Letters}\ }\textbf {\bibinfo {volume} {84}},\ \bibinfo {pages} {4685} (\bibinfo {year} {2000})}\BibitemShut {NoStop}%
\bibitem [{\citenamefont {W\'ojs}\ \emph {et~al.}(2010)\citenamefont {W\'ojs}, \citenamefont {Toke},\ and\ \citenamefont {Jain}}]{Wojs_landau_level_2010}%
  \BibitemOpen
  \bibfield  {author} {\bibinfo {author} {\bibfnamefont {A.}~\bibnamefont {W\'ojs}}, \bibinfo {author} {\bibfnamefont {C.}~\bibnamefont {Toke}}, \ and\ \bibinfo {author} {\bibfnamefont {J.~K.}\ \bibnamefont {Jain}},\ }\bibfield  {title} {\enquote {\bibinfo {title} {Landau-level mixing and the emergence of {Pfaffian} excitations for the $5/2$ fractional quantum {Hall} effect},}\ }\href {\doibase 10.1103/PhysRevLett.105.096802} {\bibfield  {journal} {\bibinfo  {journal} {Physical Review Letters}\ }\textbf {\bibinfo {volume} {105}},\ \bibinfo {pages} {096802} (\bibinfo {year} {2010})}\BibitemShut {NoStop}%
\bibitem [{\citenamefont {Storni}\ \emph {et~al.}(2010)\citenamefont {Storni}, \citenamefont {Morf},\ and\ \citenamefont {Das~Sarma}}]{Storni_fractional_2010}%
  \BibitemOpen
  \bibfield  {author} {\bibinfo {author} {\bibfnamefont {M.}~\bibnamefont {Storni}}, \bibinfo {author} {\bibfnamefont {R.~H.}\ \bibnamefont {Morf}}, \ and\ \bibinfo {author} {\bibfnamefont {S.}~\bibnamefont {Das~Sarma}},\ }\bibfield  {title} {\enquote {\bibinfo {title} {Fractional quantum {Hall} state at $\nu=$~5/2 and the {Moore}-{Read} {Pfaffian}},}\ }\href {\doibase 10.1103/PhysRevLett.104.076803} {\bibfield  {journal} {\bibinfo  {journal} {Physical Review Letters}\ }\textbf {\bibinfo {volume} {104}},\ \bibinfo {pages} {076803} (\bibinfo {year} {2010})}\BibitemShut {NoStop}%
\bibitem [{\citenamefont {Feiguin}\ \emph {et~al.}(2009)\citenamefont {Feiguin}, \citenamefont {Rezayi}, \citenamefont {Yang}, \citenamefont {Nayak},\ and\ \citenamefont {Das~Sarma}}]{Feiguin_spin_2009}%
  \BibitemOpen
  \bibfield  {author} {\bibinfo {author} {\bibfnamefont {A.~E.}\ \bibnamefont {Feiguin}}, \bibinfo {author} {\bibfnamefont {E.}~\bibnamefont {Rezayi}}, \bibinfo {author} {\bibfnamefont {K.}~\bibnamefont {Yang}}, \bibinfo {author} {\bibfnamefont {C.}~\bibnamefont {Nayak}}, \ and\ \bibinfo {author} {\bibfnamefont {S.}~\bibnamefont {Das~Sarma}},\ }\bibfield  {title} {\enquote {\bibinfo {title} {Spin polarization of the $\ensuremath{\nu}=5/2$ quantum {Hall} state},}\ }\href {\doibase 10.1103/PhysRevB.79.115322} {\bibfield  {journal} {\bibinfo  {journal} {Physical Review B}\ }\textbf {\bibinfo {volume} {79}},\ \bibinfo {pages} {115322} (\bibinfo {year} {2009})}\BibitemShut {NoStop}%
\bibitem [{\citenamefont {Peterson}\ \emph {et~al.}(2008)\citenamefont {Peterson}, \citenamefont {Jolicoeur},\ and\ \citenamefont {Das~Sarma}}]{Peterson_Finite_Layer_Thickness_2008}%
  \BibitemOpen
  \bibfield  {author} {\bibinfo {author} {\bibfnamefont {M.~R.}\ \bibnamefont {Peterson}}, \bibinfo {author} {\bibfnamefont {Th.}\ \bibnamefont {Jolicoeur}}, \ and\ \bibinfo {author} {\bibfnamefont {S.}~\bibnamefont {Das~Sarma}},\ }\bibfield  {title} {\enquote {\bibinfo {title} {Finite-layer thickness stabilizes the {Pfaffian} state for the 5/2 fractional quantum {Hall} effect: Wave function overlap and topological degeneracy},}\ }\href {\doibase 10.1103/PhysRevLett.101.016807} {\bibfield  {journal} {\bibinfo  {journal} {Physical Review Letters}\ }\textbf {\bibinfo {volume} {101}},\ \bibinfo {pages} {016807} (\bibinfo {year} {2008})}\BibitemShut {NoStop}%
\bibitem [{\citenamefont {Lee}\ \emph {et~al.}(2007)\citenamefont {Lee}, \citenamefont {Ryu}, \citenamefont {Nayak},\ and\ \citenamefont {Fisher}}]{Lee_particle_hole_2007}%
  \BibitemOpen
  \bibfield  {author} {\bibinfo {author} {\bibfnamefont {S.~S.}\ \bibnamefont {Lee}}, \bibinfo {author} {\bibfnamefont {S.}~\bibnamefont {Ryu}}, \bibinfo {author} {\bibfnamefont {C.}~\bibnamefont {Nayak}}, \ and\ \bibinfo {author} {\bibfnamefont {M.~P.~A.}\ \bibnamefont {Fisher}},\ }\bibfield  {title} {\enquote {\bibinfo {title} {Particle-hole symmetry and the $\nu=$~5/2 quantum {Hall} state},}\ }\href {\doibase 10.1103/PhysRevLett.99.236807} {\bibfield  {journal} {\bibinfo  {journal} {Physical Review Letters}\ }\textbf {\bibinfo {volume} {99}},\ \bibinfo {pages} {236807} (\bibinfo {year} {2007})}\BibitemShut {NoStop}%
\bibitem [{\citenamefont {Levin}\ \emph {et~al.}(2007)\citenamefont {Levin}, \citenamefont {Halperin},\ and\ \citenamefont {Rosenow}}]{Levin_particle_hole_2007}%
  \BibitemOpen
  \bibfield  {author} {\bibinfo {author} {\bibfnamefont {M.}~\bibnamefont {Levin}}, \bibinfo {author} {\bibfnamefont {B.~I.}\ \bibnamefont {Halperin}}, \ and\ \bibinfo {author} {\bibfnamefont {B.}~\bibnamefont {Rosenow}},\ }\bibfield  {title} {\enquote {\bibinfo {title} {Particle-hole symmetry and the {Pfaffian} state},}\ }\href {\doibase 10.1103/PhysRevLett.99.236806} {\bibfield  {journal} {\bibinfo  {journal} {Physical Review Letters}\ }\textbf {\bibinfo {volume} {99}},\ \bibinfo {pages} {236806} (\bibinfo {year} {2007})}\BibitemShut {NoStop}%
\bibitem [{\citenamefont {Rezayi}\ and\ \citenamefont {Simon}(2011)}]{Rezayi_breaking_2011}%
  \BibitemOpen
  \bibfield  {author} {\bibinfo {author} {\bibfnamefont {E.~H.}\ \bibnamefont {Rezayi}}\ and\ \bibinfo {author} {\bibfnamefont {S.~H.}\ \bibnamefont {Simon}},\ }\bibfield  {title} {\enquote {\bibinfo {title} {Breaking of particle-hole symmetry by {Landau} level mixing in the $\nu = 5/2$ quantized {Hall} state},}\ }\href {\doibase 10.1103/PhysRevLett.106.116801} {\bibfield  {journal} {\bibinfo  {journal} {Physical Review Letters}\ }\textbf {\bibinfo {volume} {106}},\ \bibinfo {pages} {116801} (\bibinfo {year} {2011})}\BibitemShut {NoStop}%
\bibitem [{\citenamefont {Pakrouski}\ \emph {et~al.}(2015)\citenamefont {Pakrouski}, \citenamefont {Peterson}, \citenamefont {Jolicoeur}, \citenamefont {Scarola}, \citenamefont {Nayak},\ and\ \citenamefont {Troyer}}]{Pakrouski_phase_2015}%
  \BibitemOpen
  \bibfield  {author} {\bibinfo {author} {\bibfnamefont {K.}~\bibnamefont {Pakrouski}}, \bibinfo {author} {\bibfnamefont {M.~R.}\ \bibnamefont {Peterson}}, \bibinfo {author} {\bibfnamefont {Th.}\ \bibnamefont {Jolicoeur}}, \bibinfo {author} {\bibfnamefont {V.~W.}\ \bibnamefont {Scarola}}, \bibinfo {author} {\bibfnamefont {C.}~\bibnamefont {Nayak}}, \ and\ \bibinfo {author} {\bibfnamefont {M.}~\bibnamefont {Troyer}},\ }\bibfield  {title} {\enquote {\bibinfo {title} {{Phase Diagram of the $\ensuremath{\nu}=5/2$ Fractional Quantum {Hall} Effect: Effects of Landau-Level Mixing and Nonzero Width}},}\ }\href {\doibase 10.1103/PhysRevX.5.021004} {\bibfield  {journal} {\bibinfo  {journal} {Physical Review X}\ }\textbf {\bibinfo {volume} {5}},\ \bibinfo {pages} {021004} (\bibinfo {year} {2015})}\BibitemShut {NoStop}%
\bibitem [{\citenamefont {Rezayi}(2017)}]{Rezayi_Landau_2017}%
  \BibitemOpen
  \bibfield  {author} {\bibinfo {author} {\bibfnamefont {E.~H.}\ \bibnamefont {Rezayi}},\ }\bibfield  {title} {\enquote {\bibinfo {title} {Landau level mixing and the ground state of the $\nu=5/2$ quantum {Hall} effect},}\ }\href {\doibase 10.1103/PhysRevLett.119.026801} {\bibfield  {journal} {\bibinfo  {journal} {Physical Review Letters}\ }\textbf {\bibinfo {volume} {119}},\ \bibinfo {pages} {026801} (\bibinfo {year} {2017})}\BibitemShut {NoStop}%
\bibitem [{\citenamefont {Banerjee}\ \emph {et~al.}(2018)\citenamefont {Banerjee}, \citenamefont {Heiblum}, \citenamefont {Umansky}, \citenamefont {Feldman}, \citenamefont {Oreg},\ and\ \citenamefont {Stern}}]{Banerjee_observation_2018}%
  \BibitemOpen
  \bibfield  {author} {\bibinfo {author} {\bibfnamefont {M.}~\bibnamefont {Banerjee}}, \bibinfo {author} {\bibfnamefont {M.}~\bibnamefont {Heiblum}}, \bibinfo {author} {\bibfnamefont {V.}~\bibnamefont {Umansky}}, \bibinfo {author} {\bibfnamefont {D.~E.}\ \bibnamefont {Feldman}}, \bibinfo {author} {\bibfnamefont {Y.}~\bibnamefont {Oreg}}, \ and\ \bibinfo {author} {\bibfnamefont {A.}~\bibnamefont {Stern}},\ }\bibfield  {title} {\enquote {\bibinfo {title} {Observation of half-integer thermal {Hall} conductance},}\ }\href {\doibase 10.1038/s41586-018-0184-1} {\bibfield  {journal} {\bibinfo  {journal} {Nature}\ }\textbf {\bibinfo {volume} {559}},\ \bibinfo {pages} {205} (\bibinfo {year} {2018})}\BibitemShut {NoStop}%
\bibitem [{\citenamefont {Dutta}\ \emph {et~al.}(2022{\natexlab{a}})\citenamefont {Dutta}, \citenamefont {Umansky}, \citenamefont {Banerjee},\ and\ \citenamefont {Heiblum}}]{Dutta_Isolated_2022}%
  \BibitemOpen
  \bibfield  {author} {\bibinfo {author} {\bibfnamefont {B.}~\bibnamefont {Dutta}}, \bibinfo {author} {\bibfnamefont {V.}~\bibnamefont {Umansky}}, \bibinfo {author} {\bibfnamefont {M.}~\bibnamefont {Banerjee}}, \ and\ \bibinfo {author} {\bibfnamefont {M.}~\bibnamefont {Heiblum}},\ }\bibfield  {title} {\enquote {\bibinfo {title} {Isolated ballistic non-{Abelian} interface channel},}\ }\href {\doibase 10.1126/science.abm6571} {\bibfield  {journal} {\bibinfo  {journal} {Science}\ }\textbf {\bibinfo {volume} {377}},\ \bibinfo {pages} {1198--1201} (\bibinfo {year} {2022}{\natexlab{a}})}\BibitemShut {NoStop}%
\bibitem [{\citenamefont {Paul}\ \emph {et~al.}(2024)\citenamefont {Paul}, \citenamefont {Tiwari}, \citenamefont {Melcer}, \citenamefont {Umansky},\ and\ \citenamefont {Heiblum}}]{Paul_Thermal_2024}%
  \BibitemOpen
  \bibfield  {author} {\bibinfo {author} {\bibfnamefont {A.~K.}\ \bibnamefont {Paul}}, \bibinfo {author} {\bibfnamefont {P.}~\bibnamefont {Tiwari}}, \bibinfo {author} {\bibfnamefont {R.~A.}\ \bibnamefont {Melcer}}, \bibinfo {author} {\bibfnamefont {V.}~\bibnamefont {Umansky}}, \ and\ \bibinfo {author} {\bibfnamefont {M.}~\bibnamefont {Heiblum}},\ }\bibfield  {title} {\enquote {\bibinfo {title} {Topological thermal hall conductance of even denominator fractional states},}\ }\href {https://arxiv.org/abs/2311.15787} {\  (\bibinfo {year} {2024})},\ \Eprint {http://arxiv.org/abs/2311.15787} {arXiv:2311.15787 [cond-mat.mes-hall]} \BibitemShut {NoStop}%
\bibitem [{\citenamefont {Melcer}\ \emph {et~al.}(2024)\citenamefont {Melcer}, \citenamefont {Gil}, \citenamefont {Paul}, \citenamefont {T.}, \citenamefont {Umansky}, \citenamefont {Heiblum}, \citenamefont {Oreg}, \citenamefont {Stern},\ and\ \citenamefont {Berg}}]{Melcer_Heat_2024}%
  \BibitemOpen
  \bibfield  {author} {\bibinfo {author} {\bibfnamefont {R.~A.}\ \bibnamefont {Melcer}}, \bibinfo {author} {\bibfnamefont {A.}~\bibnamefont {Gil}}, \bibinfo {author} {\bibfnamefont {A.~K.}\ \bibnamefont {Paul}}, \bibinfo {author} {\bibfnamefont {P.}~\bibnamefont {T.}}, \bibinfo {author} {\bibfnamefont {V.}~\bibnamefont {Umansky}}, \bibinfo {author} {\bibfnamefont {M.}~\bibnamefont {Heiblum}}, \bibinfo {author} {\bibfnamefont {Y.}~\bibnamefont {Oreg}}, \bibinfo {author} {\bibfnamefont {A.}~\bibnamefont {Stern}}, \ and\ \bibinfo {author} {\bibfnamefont {E.}~\bibnamefont {Berg}},\ }\bibfield  {title} {\enquote {\bibinfo {title} {Heat conductance of the quantum hall bulk},}\ }\href {\doibase 10.1038/s41586-023-06858-z} {\bibfield  {journal} {\bibinfo  {journal} {Nature}\ }\textbf {\bibinfo {volume} {625}},\ \bibinfo {pages} {489--493} (\bibinfo {year} {2024})}\BibitemShut {NoStop}%
\bibitem [{\citenamefont {Dutta}\ \emph {et~al.}(2022{\natexlab{b}})\citenamefont {Dutta}, \citenamefont {Yang}, \citenamefont {Melcer}, \citenamefont {Kundu}, \citenamefont {Heiblum}, \citenamefont {Umansky}, \citenamefont {Oreg}, \citenamefont {Stern},\ and\ \citenamefont {Mross}}]{Dutta_novel_2022}%
  \BibitemOpen
  \bibfield  {author} {\bibinfo {author} {\bibfnamefont {B.}~\bibnamefont {Dutta}}, \bibinfo {author} {\bibfnamefont {W.}~\bibnamefont {Yang}}, \bibinfo {author} {\bibfnamefont {R.}~\bibnamefont {Melcer}}, \bibinfo {author} {\bibfnamefont {H.~K.}\ \bibnamefont {Kundu}}, \bibinfo {author} {\bibfnamefont {M.}~\bibnamefont {Heiblum}}, \bibinfo {author} {\bibfnamefont {V.}~\bibnamefont {Umansky}}, \bibinfo {author} {\bibfnamefont {Y.}~\bibnamefont {Oreg}}, \bibinfo {author} {\bibfnamefont {A.}~\bibnamefont {Stern}}, \ and\ \bibinfo {author} {\bibfnamefont {D.}~\bibnamefont {Mross}},\ }\bibfield  {title} {\enquote {\bibinfo {title} {Distinguishing between non-{Abelian} topological orders in a quantum {Hall} system},}\ }\href {\doibase 10.1126/science.abg6116} {\bibfield  {journal} {\bibinfo  {journal} {Science}\ }\textbf {\bibinfo {volume} {375}},\ \bibinfo {pages} {193--197} (\bibinfo {year} {2022}{\natexlab{b}})}\BibitemShut {NoStop}%
\bibitem [{\citenamefont {Son}(2015)}]{Son_is_2015}%
  \BibitemOpen
  \bibfield  {author} {\bibinfo {author} {\bibfnamefont {D.~T.}\ \bibnamefont {Son}},\ }\bibfield  {title} {\enquote {\bibinfo {title} {Is the composite {Fermion} a {Dirac} particle?}}\ }\href {\doibase 10.1103/PhysRevX.5.031027} {\bibfield  {journal} {\bibinfo  {journal} {Physical Review X}\ }\textbf {\bibinfo {volume} {5}},\ \bibinfo {pages} {031027} (\bibinfo {year} {2015})}\BibitemShut {NoStop}%
\bibitem [{\citenamefont {Bonderson}\ \emph {et~al.}(2013)\citenamefont {Bonderson}, \citenamefont {Nayak},\ and\ \citenamefont {Qi}}]{Bonderson_time-reversal_2013}%
  \BibitemOpen
  \bibfield  {author} {\bibinfo {author} {\bibfnamefont {P.}~\bibnamefont {Bonderson}}, \bibinfo {author} {\bibfnamefont {C.}~\bibnamefont {Nayak}}, \ and\ \bibinfo {author} {\bibfnamefont {X.~L.}\ \bibnamefont {Qi}},\ }\bibfield  {title} {\enquote {\bibinfo {title} {A time-reversal invariant topological phase at the surface of a 3d topological insulator},}\ }\href {\doibase 10.1088/1742-5468/2013/09/P09016} {\bibfield  {journal} {\bibinfo  {journal} {J. Stat. Mech.}\ }\textbf {\bibinfo {volume} {2013}},\ \bibinfo {pages} {P09016} (\bibinfo {year} {2013})}\BibitemShut {NoStop}%
\bibitem [{\citenamefont {Chen}\ \emph {et~al.}(2014)\citenamefont {Chen}, \citenamefont {Fidkowski},\ and\ \citenamefont {Vishwanath}}]{Chen_symmetry_2014}%
  \BibitemOpen
  \bibfield  {author} {\bibinfo {author} {\bibfnamefont {X.}~\bibnamefont {Chen}}, \bibinfo {author} {\bibfnamefont {L.}~\bibnamefont {Fidkowski}}, \ and\ \bibinfo {author} {\bibfnamefont {A.}~\bibnamefont {Vishwanath}},\ }\bibfield  {title} {\enquote {\bibinfo {title} {Symmetry enforced non-{Abelian} topological order at the surface of a topological insulator},}\ }\href {\doibase 10.1103/PhysRevB.89.165132} {\bibfield  {journal} {\bibinfo  {journal} {Physical Review B}\ }\textbf {\bibinfo {volume} {89}},\ \bibinfo {pages} {165132} (\bibinfo {year} {2014})}\BibitemShut {NoStop}%
\bibitem [{\citenamefont {Levin}\ and\ \citenamefont {Halperin}(2009)}]{Levin_collective_2009}%
  \BibitemOpen
  \bibfield  {author} {\bibinfo {author} {\bibfnamefont {M.}~\bibnamefont {Levin}}\ and\ \bibinfo {author} {\bibfnamefont {B.~I}\ \bibnamefont {Halperin}},\ }\bibfield  {title} {\enquote {\bibinfo {title} {Collective states of non-{Abelian} quasiparticles in a magnetic field},}\ }\href@noop {} {\bibfield  {journal} {\bibinfo  {journal} {Physical Review B}\ }\textbf {\bibinfo {volume} {79}},\ \bibinfo {pages} {205301} (\bibinfo {year} {2009})}\BibitemShut {NoStop}%
\bibitem [{\citenamefont {Yutushui}\ \emph {et~al.}(2024{\natexlab{a}})\citenamefont {Yutushui}, \citenamefont {Hermanns},\ and\ \citenamefont {Mross}}]{Yutushui_daughters_2024}%
  \BibitemOpen
  \bibfield  {author} {\bibinfo {author} {\bibfnamefont {M.}~\bibnamefont {Yutushui}}, \bibinfo {author} {\bibfnamefont {M.}~\bibnamefont {Hermanns}}, \ and\ \bibinfo {author} {\bibfnamefont {D.~F.}\ \bibnamefont {Mross}},\ }\bibfield  {title} {\enquote {\bibinfo {title} {{Paired fermions in strong magnetic fields and daughters of even-denominator Hall plateaus}},}\ }\href {\doibase 10.1103/PhysRevB.110.165402} {\bibfield  {journal} {\bibinfo  {journal} {Physical Review B}\ }\textbf {\bibinfo {volume} {110}},\ \bibinfo {pages} {165402} (\bibinfo {year} {2024}{\natexlab{a}})}\BibitemShut {NoStop}%
\bibitem [{\citenamefont {{Zheltonozhskii}}\ \emph {et~al.}(2024)\citenamefont {{Zheltonozhskii}}, \citenamefont {{Stern}},\ and\ \citenamefont {{Lindner}}}]{Zheltonozhskii_daughters_2024}%
  \BibitemOpen
  \bibfield  {author} {\bibinfo {author} {\bibfnamefont {E.}~\bibnamefont {{Zheltonozhskii}}}, \bibinfo {author} {\bibfnamefont {A.}~\bibnamefont {{Stern}}}, \ and\ \bibinfo {author} {\bibfnamefont {N.}~\bibnamefont {{Lindner}}},\ }\bibfield  {title} {\enquote {\bibinfo {title} {{Identifying the topological order of quantized half-filled Landau levels through their daughter states}},}\ }\href@noop {} {\bibfield  {journal} {\bibinfo  {journal} {arXiv e-prints}\ } (\bibinfo {year} {2024})},\ \Eprint {http://arxiv.org/abs/2405.03780} {arXiv:2405.03780 [cond-mat.mes-hall]} \BibitemShut {NoStop}%
\bibitem [{\citenamefont {{Zhang}}\ \emph {et~al.}(2024)\citenamefont {{Zhang}}, \citenamefont {{Vishwanath}},\ and\ \citenamefont {{Wen}}}]{zhang_hierarchy_2024}%
  \BibitemOpen
  \bibfield  {author} {\bibinfo {author} {\bibfnamefont {C.}~\bibnamefont {{Zhang}}}, \bibinfo {author} {\bibfnamefont {A.}~\bibnamefont {{Vishwanath}}}, \ and\ \bibinfo {author} {\bibfnamefont {X.-G.}\ \bibnamefont {{Wen}}},\ }\bibfield  {title} {\enquote {\bibinfo {title} {{Hierarchy construction for non-abelian fractional quantum Hall states via anyon condensation}},}\ }\href@noop {} {\bibfield  {journal} {\bibinfo  {journal} {arXiv e-prints}\ } (\bibinfo {year} {2024})},\ \Eprint {http://arxiv.org/abs/2406.12068} {arXiv:2406.12068 [cond-mat.str-el]} \BibitemShut {NoStop}%
\bibitem [{\citenamefont {Manna}\ \emph {et~al.}(2022)\citenamefont {Manna}, \citenamefont {Das}, \citenamefont {Goldstein},\ and\ \citenamefont {Gefen}}]{Manna_Full_Classification_2022}%
  \BibitemOpen
  \bibfield  {author} {\bibinfo {author} {\bibfnamefont {S.}~\bibnamefont {Manna}}, \bibinfo {author} {\bibfnamefont {A.}~\bibnamefont {Das}}, \bibinfo {author} {\bibfnamefont {M.}~\bibnamefont {Goldstein}}, \ and\ \bibinfo {author} {\bibfnamefont {Y.}~\bibnamefont {Gefen}},\ }\href@noop {} {\enquote {\bibinfo {title} {Full classification of transport on an equilibrated $5/2$ edge},}\ } (\bibinfo {year} {2022}),\ \Eprint {http://arxiv.org/abs/2212.05732} {arXiv:2212.05732 [cond-mat.mes-hall]} \BibitemShut {NoStop}%
\bibitem [{\citenamefont {Yutushui}\ \emph {et~al.}(2022)\citenamefont {Yutushui}, \citenamefont {Stern},\ and\ \citenamefont {Mross}}]{Yutushui_Identifying_2022}%
  \BibitemOpen
  \bibfield  {author} {\bibinfo {author} {\bibfnamefont {M.}~\bibnamefont {Yutushui}}, \bibinfo {author} {\bibfnamefont {A.}~\bibnamefont {Stern}}, \ and\ \bibinfo {author} {\bibfnamefont {D.~F.}\ \bibnamefont {Mross}},\ }\bibfield  {title} {\enquote {\bibinfo {title} {Identifying the $\ensuremath{\nu}=\frac{5}{2}$ topological order through charge transport measurements},}\ }\href {\doibase 10.1103/PhysRevLett.128.016401} {\bibfield  {journal} {\bibinfo  {journal} {Physical Review Letters}\ }\textbf {\bibinfo {volume} {128}},\ \bibinfo {pages} {016401} (\bibinfo {year} {2022})}\BibitemShut {NoStop}%
\bibitem [{\citenamefont {Park}\ \emph {et~al.}(2024)\citenamefont {Park}, \citenamefont {Sp\aa{}nsl\"att},\ and\ \citenamefont {Mirlin}}]{Park_Fingerprints_2024}%
  \BibitemOpen
  \bibfield  {author} {\bibinfo {author} {\bibfnamefont {J.}~\bibnamefont {Park}}, \bibinfo {author} {\bibfnamefont {C.}~\bibnamefont {Sp\aa{}nsl\"att}}, \ and\ \bibinfo {author} {\bibfnamefont {A.~D.}\ \bibnamefont {Mirlin}},\ }\bibfield  {title} {\enquote {\bibinfo {title} {Fingerprints of anti-pfaffian topological order in quantum point contact transport},}\ }\href {\doibase 10.1103/PhysRevLett.132.256601} {\bibfield  {journal} {\bibinfo  {journal} {Physical Review Letters}\ }\textbf {\bibinfo {volume} {132}},\ \bibinfo {pages} {256601} (\bibinfo {year} {2024})}\BibitemShut {NoStop}%
\bibitem [{\citenamefont {Yutushui}\ and\ \citenamefont {Mross}(2023)}]{Yutushui_Identifying_2023}%
  \BibitemOpen
  \bibfield  {author} {\bibinfo {author} {\bibfnamefont {M.}~\bibnamefont {Yutushui}}\ and\ \bibinfo {author} {\bibfnamefont {D.~F.}\ \bibnamefont {Mross}},\ }\bibfield  {title} {\enquote {\bibinfo {title} {Identifying non-{Abelian} anyons with upstream noise},}\ }\href {\doibase 10.1103/PhysRevB.108.L241102} {\bibfield  {journal} {\bibinfo  {journal} {Physical Review B}\ }\textbf {\bibinfo {volume} {108}},\ \bibinfo {pages} {L241102} (\bibinfo {year} {2023})}\BibitemShut {NoStop}%
\bibitem [{\citenamefont {Protopopov}\ \emph {et~al.}(2017)\citenamefont {Protopopov}, \citenamefont {Gefen},\ and\ \citenamefont {Mirlin}}]{Protopopov_transport_2_3_2017}%
  \BibitemOpen
  \bibfield  {author} {\bibinfo {author} {\bibfnamefont {I.V.}\ \bibnamefont {Protopopov}}, \bibinfo {author} {\bibfnamefont {Y.}~\bibnamefont {Gefen}}, \ and\ \bibinfo {author} {\bibfnamefont {A.D.}\ \bibnamefont {Mirlin}},\ }\bibfield  {title} {\enquote {\bibinfo {title} {Transport in a disordered $\nu=2/3$ fractional quantum {Hall} junction},}\ }\href {\doibase https://doi.org/10.1016/j.aop.2017.07.015} {\bibfield  {journal} {\bibinfo  {journal} {Annals of Physics}\ }\textbf {\bibinfo {volume} {385}},\ \bibinfo {pages} {287 -- 327} (\bibinfo {year} {2017})}\BibitemShut {NoStop}%
\bibitem [{\citenamefont {Haldane}(1995)}]{Haldane_stability_1995}%
  \BibitemOpen
  \bibfield  {author} {\bibinfo {author} {\bibfnamefont {F.~D.~M.}\ \bibnamefont {Haldane}},\ }\bibfield  {title} {\enquote {\bibinfo {title} {Stability of chiral {Luttinger} liquids and abelian quantum {Hall} states},}\ }\href {\doibase 10.1103/PhysRevLett.74.2090} {\bibfield  {journal} {\bibinfo  {journal} {Physical Review Letters}\ }\textbf {\bibinfo {volume} {74}},\ \bibinfo {pages} {2090--2093} (\bibinfo {year} {1995})}\BibitemShut {NoStop}%
\bibitem [{\citenamefont {Wen}(1991{\natexlab{b}})}]{Wen_edge_1991}%
  \BibitemOpen
  \bibfield  {author} {\bibinfo {author} {\bibfnamefont {X.~G.}\ \bibnamefont {Wen}},\ }\bibfield  {title} {\enquote {\bibinfo {title} {Edge transport properties of the fractional quantum {Hall} states and weak-impurity scattering of a one-dimensional charge-density wave},}\ }\href {\doibase 10.1103/PhysRevB.44.5708} {\bibfield  {journal} {\bibinfo  {journal} {Physical Review B}\ }\textbf {\bibinfo {volume} {44}},\ \bibinfo {pages} {5708--5719} (\bibinfo {year} {1991}{\natexlab{b}})}\BibitemShut {NoStop}%
\bibitem [{\citenamefont {Moore}\ and\ \citenamefont {Wen}(1998)}]{Moore_Classification_1997}%
  \BibitemOpen
  \bibfield  {author} {\bibinfo {author} {\bibfnamefont {J.~E.}\ \bibnamefont {Moore}}\ and\ \bibinfo {author} {\bibfnamefont {X.~G.}\ \bibnamefont {Wen}},\ }\bibfield  {title} {\enquote {\bibinfo {title} {Classification of disordered phases of quantum {Hall} edge states},}\ }\href {\doibase 10.1103/PhysRevB.57.10138} {\bibfield  {journal} {\bibinfo  {journal} {Physical Review B}\ }\textbf {\bibinfo {volume} {57}},\ \bibinfo {pages} {10138--10156} (\bibinfo {year} {1998})}\BibitemShut {NoStop}%
\bibitem [{\citenamefont {Bishara}\ \emph {et~al.}(2008)\citenamefont {Bishara}, \citenamefont {Fiete},\ and\ \citenamefont {Nayak}}]{Bishara_PH_Read_Rezayi_2008}%
  \BibitemOpen
  \bibfield  {author} {\bibinfo {author} {\bibfnamefont {W.}~\bibnamefont {Bishara}}, \bibinfo {author} {\bibfnamefont {G.~A.}\ \bibnamefont {Fiete}}, \ and\ \bibinfo {author} {\bibfnamefont {C.}~\bibnamefont {Nayak}},\ }\bibfield  {title} {\enquote {\bibinfo {title} {Quantum {Hall} states at $\ensuremath{\nu}=\frac{2}{k+2}$: Analysis of the particle-hole conjugates of the general level-$k$ read-rezayi states},}\ }\href {\doibase 10.1103/PhysRevB.77.241306} {\bibfield  {journal} {\bibinfo  {journal} {Physical Review B}\ }\textbf {\bibinfo {volume} {77}},\ \bibinfo {pages} {241306(R)} (\bibinfo {year} {2008})}\BibitemShut {NoStop}%
\bibitem [{\citenamefont {Kane}\ \emph {et~al.}(1994)\citenamefont {Kane}, \citenamefont {Fisher},\ and\ \citenamefont {Polchinski}}]{Kane_Randomness_1994}%
  \BibitemOpen
  \bibfield  {author} {\bibinfo {author} {\bibfnamefont {C.~L.}\ \bibnamefont {Kane}}, \bibinfo {author} {\bibfnamefont {M.~P.~A.}\ \bibnamefont {Fisher}}, \ and\ \bibinfo {author} {\bibfnamefont {J.}~\bibnamefont {Polchinski}},\ }\bibfield  {title} {\enquote {\bibinfo {title} {Randomness at the edge: Theory of quantum {Hall} transport at filling \ensuremath{\nu}=2/3},}\ }\href {\doibase 10.1103/PhysRevLett.72.4129} {\bibfield  {journal} {\bibinfo  {journal} {Physical Review Letters}\ }\textbf {\bibinfo {volume} {72}},\ \bibinfo {pages} {4129--4132} (\bibinfo {year} {1994})}\BibitemShut {NoStop}%
\bibitem [{Note1()}]{Note1}%
  \BibitemOpen
  \bibinfo {note} {For $|\ell |\geq 2$, the Majorana modes form complex fermions $\gamma _{2k-1}+i\gamma _{2k}\propto e^{i\phi _k}$ that could carry $U(1)$ current $I_{k}=i\partial _\tau \phi _k/2\pi $ (not to be confused with the thermal current). However, these currents are not protected by charge conservation and dissipate in the presence of disorder. Thus, the impinging neutral $U(1)$ currents on the constriction are zero at $x_R$($x_L$) for $\ell \geq 2$($\ell \leq -2$). If these currents were non-zero $\DOTSB \sum@ \slimits@ _k I_k=I_{U(1)}$, this would result in induced charge currents $I_\protect \text {D1}=-I_\protect \text {D2}\propto I_{U(1)}$ for source voltage $V=0$.}\BibitemShut {Stop}%
\bibitem [{\citenamefont {Kurilovich}\ \emph {et~al.}(2023)\citenamefont {Kurilovich}, \citenamefont {Raines},\ and\ \citenamefont {Glazman}}]{Kurilovich_Disorder_2023}%
  \BibitemOpen
  \bibfield  {author} {\bibinfo {author} {\bibfnamefont {V.~D.}\ \bibnamefont {Kurilovich}}, \bibinfo {author} {\bibfnamefont {Z.~M.}\ \bibnamefont {Raines}}, \ and\ \bibinfo {author} {\bibfnamefont {L.~I.}\ \bibnamefont {Glazman}},\ }\bibfield  {title} {\enquote {\bibinfo {title} {{Disorder-enabled Andreev reflection of a quantum Hall edge}},}\ }\href {\doibase 10.1038/s41467-023-37794-1} {\bibfield  {journal} {\bibinfo  {journal} {Nature Communications}\ }\textbf {\bibinfo {volume} {14}},\ \bibinfo {pages} {2237} (\bibinfo {year} {2023})}\BibitemShut {NoStop}%
\bibitem [{\citenamefont {Lichtman}\ \emph {et~al.}(2021)\citenamefont {Lichtman}, \citenamefont {Thorngren}, \citenamefont {Lindner}, \citenamefont {Stern},\ and\ \citenamefont {Berg}}]{Lichtman_bulk_anyons_2021}%
  \BibitemOpen
  \bibfield  {author} {\bibinfo {author} {\bibfnamefont {T.}~\bibnamefont {Lichtman}}, \bibinfo {author} {\bibfnamefont {R.}~\bibnamefont {Thorngren}}, \bibinfo {author} {\bibfnamefont {N.~H.}\ \bibnamefont {Lindner}}, \bibinfo {author} {\bibfnamefont {A.}~\bibnamefont {Stern}}, \ and\ \bibinfo {author} {\bibfnamefont {E.}~\bibnamefont {Berg}},\ }\bibfield  {title} {\enquote {\bibinfo {title} {Bulk anyons as edge symmetries: Boundary phase diagrams of topologically ordered states},}\ }\href {\doibase 10.1103/PhysRevB.104.075141} {\bibfield  {journal} {\bibinfo  {journal} {Physical Review B}\ }\textbf {\bibinfo {volume} {104}},\ \bibinfo {pages} {075141} (\bibinfo {year} {2021})}\BibitemShut {NoStop}%
\bibitem [{\citenamefont {Di~Francesco}\ \emph {et~al.}(1997)\citenamefont {Di~Francesco}, \citenamefont {Mathieu},\ and\ \citenamefont {Sénéchal}}]{DiFrancesco_CFT_1997}%
  \BibitemOpen
  \bibfield  {author} {\bibinfo {author} {\bibfnamefont {P.}~\bibnamefont {Di~Francesco}}, \bibinfo {author} {\bibfnamefont {P.}~\bibnamefont {Mathieu}}, \ and\ \bibinfo {author} {\bibfnamefont {D.}~\bibnamefont {Sénéchal}},\ }\href {\doibase 10.1007/978-1-4612-2256-9} {\emph {\bibinfo {title} {Conformal Field Theory}}},\ Graduate Texts in Contemporary Physics\ (\bibinfo  {publisher} {Springer},\ \bibinfo {address} {New York, NY},\ \bibinfo {year} {1997})\BibitemShut {NoStop}%
\bibitem [{\citenamefont {Grosfeld}\ and\ \citenamefont {Schoutens}(2009)}]{Grosfeld_tricritial_2009}%
  \BibitemOpen
  \bibfield  {author} {\bibinfo {author} {\bibfnamefont {E.}~\bibnamefont {Grosfeld}}\ and\ \bibinfo {author} {\bibfnamefont {K.}~\bibnamefont {Schoutens}},\ }\bibfield  {title} {\enquote {\bibinfo {title} {Non-abelian anyons: When ising meets fibonacci},}\ }\href {\doibase 10.1103/PhysRevLett.103.076803} {\bibfield  {journal} {\bibinfo  {journal} {Physical Review Letters}\ }\textbf {\bibinfo {volume} {103}},\ \bibinfo {pages} {076803} (\bibinfo {year} {2009})}\BibitemShut {NoStop}%
\bibitem [{\citenamefont {Maslov}\ and\ \citenamefont {Stone}(1995)}]{Maslov_Landauer_conductance_1995}%
  \BibitemOpen
  \bibfield  {author} {\bibinfo {author} {\bibfnamefont {D.~L.}\ \bibnamefont {Maslov}}\ and\ \bibinfo {author} {\bibfnamefont {M.}~\bibnamefont {Stone}},\ }\bibfield  {title} {\enquote {\bibinfo {title} {Landauer conductance of { {Luttinger} } liquids with leads},}\ }\href {\doibase 10.1103/PhysRevB.52.R5539} {\bibfield  {journal} {\bibinfo  {journal} {Physical Review B}\ }\textbf {\bibinfo {volume} {52}},\ \bibinfo {pages} {R5539--R5542} (\bibinfo {year} {1995})}\BibitemShut {NoStop}%
\bibitem [{\citenamefont {Rosenow}\ and\ \citenamefont {Halperin}(2010)}]{Rosenow_signatures_2010}%
  \BibitemOpen
  \bibfield  {author} {\bibinfo {author} {\bibfnamefont {B.}~\bibnamefont {Rosenow}}\ and\ \bibinfo {author} {\bibfnamefont {B.~I.}\ \bibnamefont {Halperin}},\ }\bibfield  {title} {\enquote {\bibinfo {title} {Signatures of neutral quantum {Hall} modes in transport through low-density constrictions},}\ }\href {\doibase 10.1103/PhysRevB.81.165313} {\bibfield  {journal} {\bibinfo  {journal} {Physical Review B}\ }\textbf {\bibinfo {volume} {81}},\ \bibinfo {pages} {165313} (\bibinfo {year} {2010})}\BibitemShut {NoStop}%
\bibitem [{\citenamefont {Yutushui}\ \emph {et~al.}(2024{\natexlab{b}})\citenamefont {Yutushui}, \citenamefont {Park},\ and\ \citenamefont {Mirlin}}]{Yutushui_Localization_2024}%
  \BibitemOpen
  \bibfield  {author} {\bibinfo {author} {\bibfnamefont {M.}~\bibnamefont {Yutushui}}, \bibinfo {author} {\bibfnamefont {J.}~\bibnamefont {Park}}, \ and\ \bibinfo {author} {\bibfnamefont {A.~D.}\ \bibnamefont {Mirlin}},\ }\bibfield  {title} {\enquote {\bibinfo {title} {{Localization and conductance in fractional quantum Hall edges}},}\ }\href {\doibase 10.1103/PhysRevB.110.035402} {\bibfield  {journal} {\bibinfo  {journal} {Physical Review B}\ }\textbf {\bibinfo {volume} {110}},\ \bibinfo {pages} {035402} (\bibinfo {year} {2024}{\natexlab{b}})}\BibitemShut {NoStop}%
\bibitem [{\citenamefont {Read}\ and\ \citenamefont {Rezayi}(1999)}]{Read_Beyond_1999}%
  \BibitemOpen
  \bibfield  {author} {\bibinfo {author} {\bibfnamefont {N.}~\bibnamefont {Read}}\ and\ \bibinfo {author} {\bibfnamefont {E.}~\bibnamefont {Rezayi}},\ }\bibfield  {title} {\enquote {\bibinfo {title} {Beyond paired quantum {Hall} states: Parafermions and incompressible states in the first excited {Landau} level},}\ }\href {\doibase 10.1103/PhysRevB.59.8084} {\bibfield  {journal} {\bibinfo  {journal} {Physical Review B}\ }\textbf {\bibinfo {volume} {59}},\ \bibinfo {pages} {8084--8092} (\bibinfo {year} {1999})}\BibitemShut {NoStop}%
\bibitem [{\citenamefont {{Zamolodchikov}}\ and\ \citenamefont {{Fateev}}(1985)}]{Zamolodchikov_parafermion_1985}%
  \BibitemOpen
  \bibfield  {author} {\bibinfo {author} {\bibfnamefont {A.~B.}\ \bibnamefont {{Zamolodchikov}}}\ and\ \bibinfo {author} {\bibfnamefont {V.~A.}\ \bibnamefont {{Fateev}}},\ }\bibfield  {title} {\enquote {\bibinfo {title} {{Nonlocal (parafermion) currents in two-dimensional conformal quantum field theory and self-dual critical points in ZN-symmetric statistical systems}},}\ }\href@noop {} {\bibfield  {journal} {\bibinfo  {journal} {Soviet Journal of Experimental and Theoretical Physics}\ }\textbf {\bibinfo {volume} {62}},\ \bibinfo {pages} {215} (\bibinfo {year} {1985})}\BibitemShut {NoStop}%
\bibitem [{\citenamefont {{Karabali}}\ and\ \citenamefont {{Schnitzer}}(1990)}]{Karabali_WZW_1990}%
  \BibitemOpen
  \bibfield  {author} {\bibinfo {author} {\bibfnamefont {D.}~\bibnamefont {{Karabali}}}\ and\ \bibinfo {author} {\bibfnamefont {H.~J.}\ \bibnamefont {{Schnitzer}}},\ }\bibfield  {title} {\enquote {\bibinfo {title} {{BRST quantization of the gauged WZW action and coset conformal field theories}},}\ }\href {\doibase 10.1016/0550-3213(90)90075-O} {\bibfield  {journal} {\bibinfo  {journal} {Nuclear Physics B}\ }\textbf {\bibinfo {volume} {329}},\ \bibinfo {pages} {649--666} (\bibinfo {year} {1990})}\BibitemShut {NoStop}%
\bibitem [{\citenamefont {{Witten}}(1984)}]{Witten_non_abelian_bosonization_1984}%
  \BibitemOpen
  \bibfield  {author} {\bibinfo {author} {\bibfnamefont {E.}~\bibnamefont {{Witten}}},\ }\bibfield  {title} {\enquote {\bibinfo {title} {{Non-abelian bosonization in two dimensions}},}\ }\href {\doibase 10.1007/BF01215276} {\bibfield  {journal} {\bibinfo  {journal} {Communications in Mathematical Physics}\ }\textbf {\bibinfo {volume} {92}},\ \bibinfo {pages} {455--472} (\bibinfo {year} {1984})}\BibitemShut {NoStop}%
\end{thebibliography}%
\end{document}